%% file: main.tex
\documentclass[conference]{IEEEtran}
\IEEEoverridecommandlockouts
\usepackage{cite}
\usepackage{amsmath,amssymb,amsfonts}
\usepackage{algorithmic}
\usepackage{dblfloatfix}
\usepackage{tabularx}
\usepackage{float}
\usepackage{graphicx}
\usepackage{textcomp}
\usepackage{xcolor}
\usepackage{multirow}
\usepackage{array}
\usepackage{enumitem}
\usepackage{subcaption}
\usepackage{float}
\usepackage{xspace}
\usepackage{hyperref}
\usepackage{booktabs}
\usepackage{tcolorbox}
\def\BibTeX{{\rm B\kern-.05em{\sc i\kern-.025em b}\kern-.08em
    T\kern-.1667em\lower.7ex\hbox{E}\kern-.125emX}}
\begin{document}



\newcommand{\temp}[1]{{\color{red}#1}}
\newcommand{\tempfig}{{\color{red}xx}\xspace}
\newcommand{\divider}{{\color{red}\noindent\rule{\textwidth}{0.3pt}\newline}}
\newcommand{\bolderandunderline}[1]{\textbf{\underline{#1}}}

\newcommand{\stackoverflow}{Stack Overflow\xspace}
\newcommand{\javascript}{JavaScript\xspace}
\newcommand{\figurex}{{\colorbox{yellow}{\color{red}xx\xspace}}}
\newcommand{\tradeoff}{{trade-off\xspace}}
\newcommand{\authorplusetal}[1]{#1 et al.\xspace}
\newcommand{\needcite}{{\color{red}[cite]}\xspace}
\newcommand{\citethese}[1]{{\color{orange}[Cite: #1]}\xspace}
\newcommand{\code}[1]{\texttt{#1}}
\newcommand{\userquote}[1]{``\textit{#1}''}

\newcommand{\gist}[1]{{\color{byzantine}$\boldsymbol{\cdot}$ #1}}

\newcommand{\llms}{LLMs\xspace}
\newcommand{\chatgpt}{\texttt{ChatGPT}\xspace}
\newcommand{\gptThreePointFive}{\texttt{GPT-3.5}\xspace}
\newcommand{\gptThreePointFiveAPI}{\texttt{gpt-3.5-turbo}\xspace}
\newcommand{\gptFour}{\texttt{GPT-4}\xspace}
\newcommand{\gptTwo}{\texttt{GPT-2}\xspace}

\newcommand{\lxieyangstoppedhere}{{\color{blue}\vspace{.1cm}\textbf{<<<lxieyang stopped here>>>} \vspace{.1cm}}}

\newcommand{\systemname}{MobileMaker\xspace} 
\newcommand{\mobilemaker}{MobileMaker\xspace}
\newcommand{\prototypeVideoIdeas}{Img2VideoIdeas\xspace}
\newcommand{\prototypePlaylist}{Img2Playlist\xspace}

\newcommand{\savvas}[1]{{\color{black}#1}}
\newcommand{\michael}[1]{{\color{black}#1}}

\title{\textit{In Situ} AI Prototyping: Infusing Multimodal Prompts into Mobile Settings with \systemname 
\vspace{-3mm}}

\author{
\IEEEauthorblockN{
Savvas Petridis$^*$, Michael Xieyang Liu$^*$, Alexander J. Fiannaca, Vivian Tsai, Michael Terry, Carrie J. Cai
}
\IEEEauthorblockA{
Google DeepMind, USA
\\ \{petridis,lxieyang,afiannaca,vivtsai,michaelterry,cjcai\}@google.com
}
}


\maketitle

\begin{abstract}
Recent advances in multimodal large language models (LLMs) have made it easier to rapidly prototype AI-powered features, especially for mobile use cases. 
However, gathering early, mobile-situated user feedback on these AI prototypes remains challenging.
The broad scope and flexibility of LLMs means that, for a given use-case-specific prototype, there is a crucial need to understand the wide range of in-the-wild input users are likely to provide and their in-context expectations for the AI's behavior. 
To explore the concept of \textit{in situ} AI prototyping and testing, we created \systemname: a platform that enables designers to rapidly create and test mobile AI prototypes directly on devices. This tool also enables testers to make on-device, in-the-field revisions of prototypes using natural language. In an exploratory study with 16 participants, we explored how user feedback on prototypes created with \systemname compares to that of existing prototyping tools (e.g., Figma, prompt editors). Our findings suggest that \systemname prototypes enabled more serendipitous discovery of: model input edge cases, discrepancies between AI's and user's in-context interpretation of the task, and contextual signals missed by the AI. Furthermore, we learned that while the ability to make in-the-wild revisions led users to feel more fulfilled as active participants in the design process, it might also constrain their feedback to the subset of changes perceived as more actionable or implementable by the prototyping tool.\looseness=-1


%

\end{abstract}

\begin{IEEEkeywords}
Prototyping, LLMs, Generative AI, Design
\end{IEEEkeywords}

{\def\thefootnote{}\footnotetext{$^*$Equal contribution.}}

\input{sections/sec-010-intro}

\input{sections/sec-020-related-work}

\input{sections/sec-030-formative-study}

\input{sections/sec-040-system}

\input{sections/sec-050-study}

\input{sections/sec-060-findings}

\input{sections/sec-070-design-reflection}

\input{sections/sec-080-discussion}

\input{sections/sec-090-conclusion}


\bibliographystyle{IEEEtran}
\bibliography{references, additional}

\clearpage
\appendix
\setcounter{figure}{0}
\renewcommand{\thefigure}{A\arabic{figure}} 
\input{appendix/prompt-creation}
\input{appendix/prompt-classifier}
\input{appendix/prompt-revision}
\input{appendix/structure-revision}

\end{document}

%% file: sections/sec-010-intro.tex
\section{Introduction}

Recent advances in LLMs have lowered the barriers to rapidly prototyping novel AI-powered features and products via prompt programming \cite{jiang_promptmaker_2022,wu_promptchainer_2022,wu_ai_2022,liu_selenite_2023}. Compared to the traditional, more time-consuming AI development cycle of data collection, model training, and integration into UIs \cite{yang_investigating_2018,yang_re-examining_2020}, it is now much easier to prototype AI through ``prompt-based prototyping'' \cite{petridis_promptinfuser_2023-1,petridis_promptinfuser_2023,liu_we_2024}: crafting LLM prompts in minutes and embedding them into new AI prototypes. Simultaneously, the multimodality of recent LLMs (e.g., Gemini \cite{team_gemini_2023}, GPT4 \cite{openai_gpt-4_2023}) has made it possible to process rich inputs (e.g., text, images, video), unlocking a wide range of use cases suitable for mobile or ``on the go'' settings, from diagnosing plant diseases (from a photo and text list of symptoms) to generating recipes (based on a picture of ingredients and set of dietary restrictions).

\michael{Despite rapid advancement of LLMs and AI prototyping, gathering \textit{early \michael{tester} feedback} on AI prototypes remains challenging, especially in mobile environments}. Previous research on ``embodied'' user feedback \cite{dourish_where_2004, rogers_why_2007, crabtree_introduction_2013} suggests that real-world, in-the-wild settings \michael{yield} more authentic \michael{tester} feedback than controlled lab settings, \michael{as prototypes are used within their intended context}. 
\michael{This is especially crucial for LLM-powered AI prototypes due to LLMs' specific idiosyncrasies and flexibility.}
\michael{Unlike traditional machine learning models, which are typically scoped to specific, narrow applications with inputs and outputs relatively known a priori \cite{harrison_skinput_2010, hudson_predicting_2003, langley_machine_1997}, LLMs can be applied to a broad spectrum of tasks (e.g., code completion, medical help, tutoring), receive diverse ``natural'' inputs (e.g., photography, icons, diagrams, text, audio), and function in various real-world settings (e.g., home, commute, work, coffee shop).}
Given this broad scope, for a given use case-specific prototype, there is the need to understand the types of \textit{input} \michael{testers are} likely to provide, their \michael{context-specific} \textit{expectations} of the AI's behavior, and validation of the overall prototype \textit{concept}.

\input{figures/figure-1}

LLMs also open up new opportunities for in situ prototyping and testing. First, because prompt programming has significantly reduced the ML expertise required to prototype AI, there is an opportunity for designers to more easily bring LLM-powered prototypes to life within mobile user interfaces (UIs) for in-the-wild experiences. Secondly, much as paper prototyping enables \michael{designers and testers}\footnote{In this paper, we define ``designers'' as those who create prototypes, and ``testers'' as individuals who use these prototypes and provide user feedback.} to alter designs on-the-fly, LLM-powered prototypes could enable testers to proactively revise and improve prototypes \textit{during} testing, at the very moments they encounter an issue or get an idea for an improvement to the design.\looseness=-1

To explore these opportunities for \textbf{in situ AI prototyping and testing}, we created \systemname: an AI prototyping tool that \michael{1) enables designers to rapidly create a mobile AI prototype that can be tested on-device, and 2) enables testers to revise the prototype's design in-the-wild simply through natural language (NL).} Specifically, \systemname allows \michael{testers} to provide not only feedback, but also alternative, functional prototypes that better meet their specific needs \michael{and can be instantly re-tested}. 
\michael{We conducted an exploratory study to examine how tester feedback using \systemname (on functional LLM-powered prototypes) compares to and complements traditional user-feedback mechanisms, and to assess the impact of in-the-wild revisions on the testing experience.} In a subsequent reflection activity, 
two professional designers discussed how the \michael{tester-generated} revisions could influence the iterative design process.\looseness=-1

We found that \systemname prototypes enabled \michael{testers} to authentically experience the envisioned artifact, which allowed them to 1) evaluate if the overall concept was mobile appropriate, 2) experiment with serendipitous edge cases via unconventional, in-the-wild inputs, 3) discover discrepancies between their interpretation of the task and the AI's interpretation, and 4) more critically evaluate the AI's output using the richer contextual cues in their surroundings. 
We also discovered that the ability to make \michael{on-the-spot} prototype revisions led \michael{testers} to feel more \michael{engaged} as \textit{active participants} in the design process and more critically assess their own feedback. 
Furthermore, though designers were initially hesitant to let \michael{testers} ``short-circuit'' the design process, they \michael{later recognized} revised prototypes as a more powerful and convincing form of feedback, especially because they felt traditional user feedback could be more easily overlooked or dismissed. 
In sum, this paper makes the following contributions:

\begin{itemize}[leftmargin=0.15in]
\item Design goals for LLM-powered mobile AI prototyping platforms, based on formative studies and iterative feedback from designers,

\item \systemname, a prototyping tool that enables designers to create, test, and revise AI prototypes in NL while on mobile,

\item an exploratory study showing how in situ prototypes (created using \systemname) led to qualitatively different \michael{tester} feedback compared to that of traditional AI prototypes,\looseness=-1 

\item a follow-up reflection activity with two designers, showing how designers can make use of fully functional revised prototypes created by \michael{testers} in the wild,

\item a discussion of implications for future design and prototyping platforms, advocating for systems that prioritize quick iterations and authentic \michael{tester} feedback for AI experienced in-the-wild.

\end{itemize}

%% file: figures/figure-1.tex
\begin{figure}[t]
\centering
    \begin{subfigure}[t]{0.48\linewidth}
    \centering
    \includegraphics[height=1.4in]{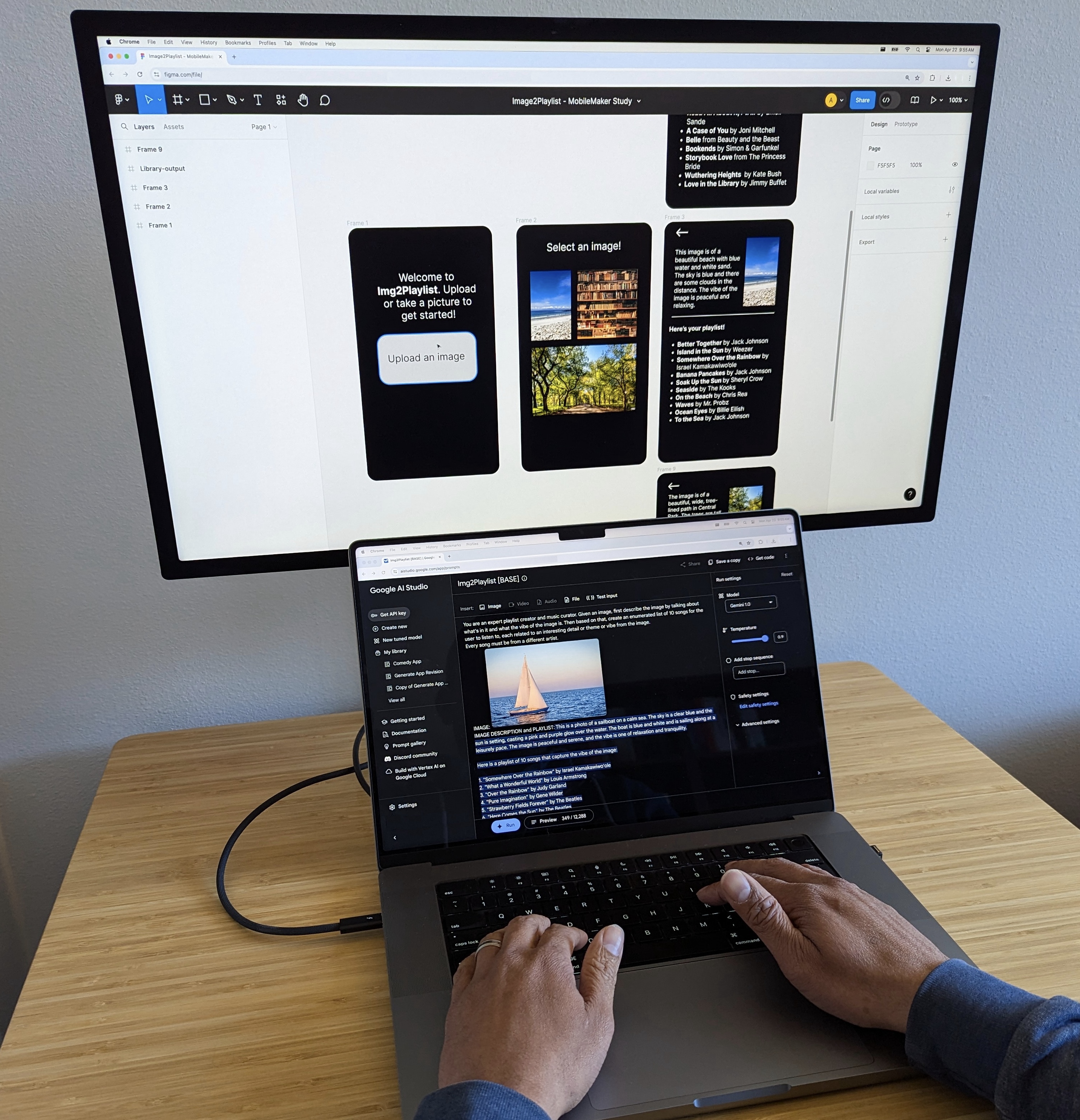}
    \vspace{-1mm}
    \caption{\footnotesize Desktop prototyping and testing\label{traditional}}
    \end{subfigure}
    ~
    \begin{subfigure}[t]{0.48\linewidth}
    \centering
    \includegraphics[height=1.4in]{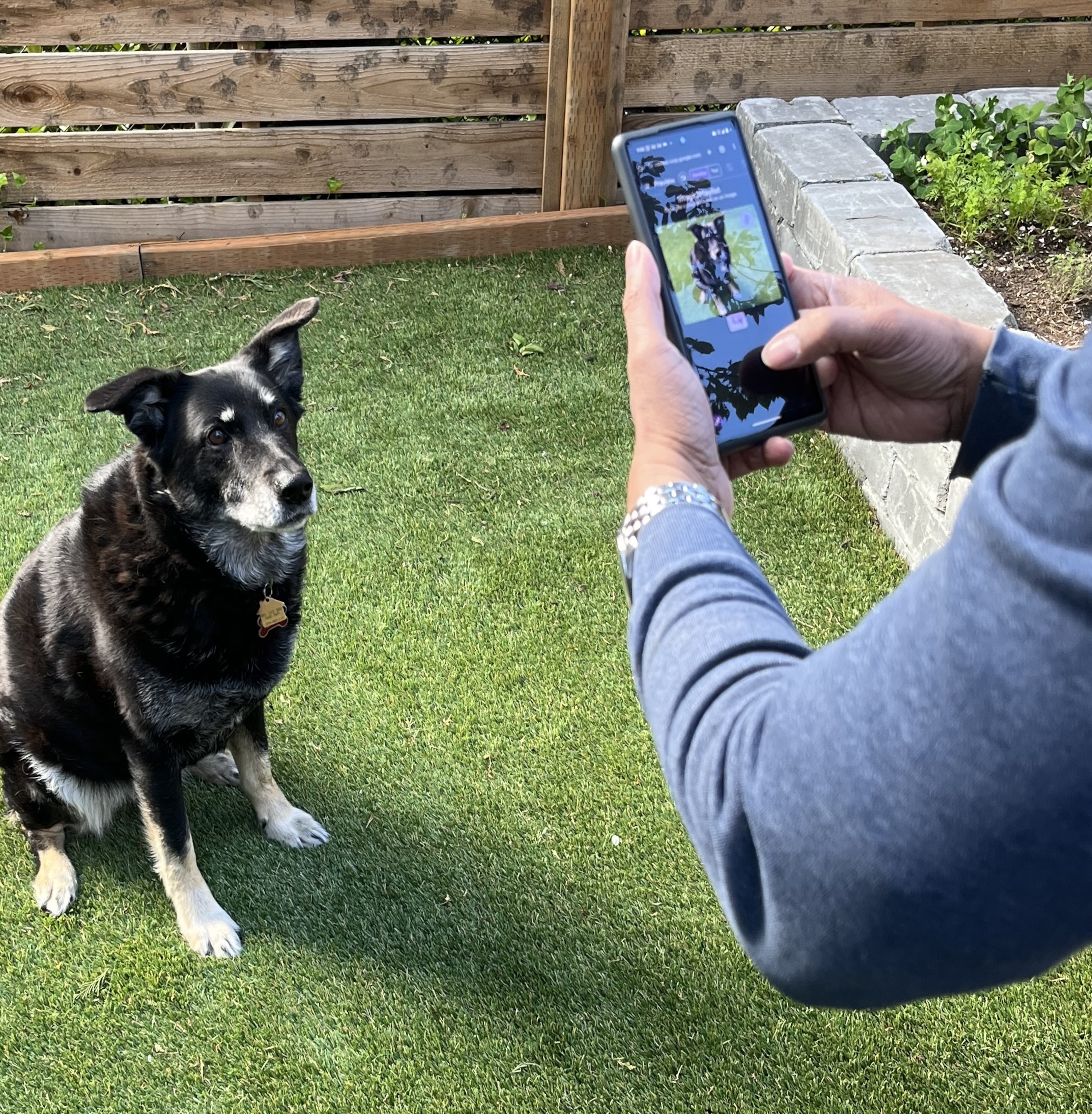}
    \vspace{-1mm}
    \caption{\footnotesize In situ prototyping and testing\label{insitu}}
    \end{subfigure}
\vspace{-1mm}
\caption{Testing AI prototypes on desktop, e.g. using UI mockups and LLM prompt editors (\subref{traditional}) vs. on mobile (\subref{insitu}): \systemname helps designers quickly get functional AI prototypes onto mobile devices and experienced in the wild for early feedback, and enables testers to revise and re-configure the AI prototype on-the-fly, while in the field.\looseness=-1}
\label{fig:teaser}
\vspace{-6mm}
\end{figure}

%% file: sections/sec-020-related-work.tex
\section{Related Work}\label{sec:rw}

\subsection{Prototyping with AI}

Prototyping is a fundamental step that allows designers to explore and refine product ideas and user interfaces \cite{beaudouin-lafon_prototyping_2007,lim_anatomy_2008}. Prototypes can range from low-fidelity \cite{sefelin_paper_2003,snyder_paper_2003}, which might just sketch out user flows \cite{de_sa_mobile_2009}, to high-fidelity, which closely mimic the final product in both look and functionality \cite{rudd_low_1996,walker_high-fidelity_2002}. However, integrating AI into these prototypes has been notoriously challenging--designers often struggle to grasp the capabilities and limitations of AI \cite{dove_ux_2017,yang_grounding_2018,yang_mapping_2018,liu_crystalline_2022},
and building functional prototypes of AI requires substantial ML expertise and engineering effort \cite{girardin_when_2017,yang_sketching_2019,kayacik_identifying_2019}.\looseness=-1

Recent advances in LLMs have dramatically reduced the barriers to prototyping AI functionality \cite{jiang_promptmaker_2022,petridis_constitutionmaker_2023,liu_what_2023,chung_talebrush_2022,wu_promptchainer_2022} through natural language prompting \cite{brown_language_2020,ouyang_training_2022,kahng_llm_2024,wu_ai_2022,petridis_constitutionalexperts_2024}. 
In addition, systems like PromptInfuser \cite{petridis_promptinfuser_2023-1,petridis_promptinfuser_2023} and ProtoAI \cite{subramonyam_protoai_2021} have been developed to help designers prototype AI functionalities within the context of UI, enabling them to produce prototypes that realistically represent the envisioned artifact and better anticipating UI issues and technical constraints \cite{petridis_promptinfuser_2023}. 
\systemname extends this capability by wrapping LLM-powered prompts and their outputs ``in an app UI shell,'' with a particular focus on mobile scenarios. However, unlike previous approaches that focus on desktop-bound prototyping \cite{petridis_promptinfuser_2023,feng_canvil_2024}, prototypes created with \systemname can be used on actual mobile devices, offering more authentic user experiences and interactions. Furthermore, \systemname supports transforming natural language ideas directly into working prototypes, reducing the initial barrier of building and configuring prototypes from scratch as well as the potential learning curve associated with onboarding to a new system.\looseness=-1

\subsection{In Situ User Testing}
After developing a prototype, it is crucial to perform user testing and gather feedback \cite{walker_high-fidelity_2002,boothe_effects_2013}. Traditional UI prototyping tools like Figma \cite{figma_figma_2024} allow \michael{testers} to interact with click-through prototypes \cite{figma_figma_2024,balsamiq_balsamiq_2024,sketch_sketch_2024}, but these often include only pre-defined input and output examples \cite{sefelin_paper_2003,feng_addressing_2023}, and the quality and relevance of these examples may not fully represent the diverse contexts and challenges that end-users may face in the real-world \cite{yang_machine_2018}. 
To circumvent the technical demands of building functional AI prototypes while still engaging end-users, designers often conduct user testing with Wizard of Oz setups \cite{maulsby_prototyping_1993}, where a human simulates AI responses \cite{cranshaw_calendar.help:_2017,klemmer_suede_2000,riek_wizard_2012}. Although helpful, this often fails to accurately simulate the unique errors and behaviors of actual AI systems \cite{parnin_building_2023,kulkarni_word_2023}.\looseness=-1

To address these challenges, in this work, we focus on supporting ``in situ'' user testing of AI prototypes, i.e., enabling interactions with \systemname prototypes in their intended real-world environments \cite{crabtree_introduction_2013,rogers_why_2007,nogueira_effectiveness_2023,dzvonyar_context-aware_2016}. Specifically, \systemname allows users to capture inputs from their immediate surroundings using their mobile devices and view AI model outputs in that same context. We found that this approach not only facilitated the discovery of serendipitous edge cases from real-world inputs (echoing findings from \cite{tseng_co-ml_2024,dwivedi_exploring_2021} in the context of in situ ML model testing), but also enabled \michael{testers} to evaluate the AI's performance more critically, leveraging the rich contextual cues around them.

\subsection{Iterative Design Based on User Feedback}

In HCI, iterative design is recognized as a critical method that emphasizes the repeated cycle of designing, testing, and refining products based on user feedback \cite{buxton_iteration_1980,dow_wizard_2005,nielsen_iterative_1993}, 
which is crucial in helping designers identify user pain points and areas for improvements \cite{dow_wizard_2005,knapp_sprint_2016}. The length of each iteration in the design process can vary significantly--ranging from a few days in rapid prototyping scenarios to several weeks or months in more sophisticated systems--based on factors like the iteration's specific objectives, the level of fidelity of the prototype involved, and the available resources \cite{nogueira_effectiveness_2023,norman_design_2013,ulrich_product_2016,brown_change_2011}.
In this work, we explore the opportunity to accelerate these iteration cycles even further by having \systemname instantly implement users' feedback on the spot, thereby enabling them to immediately interact with a revised version of the prototype. This allowed our study participants to more critically reflect on their own feedback and increased their perceived engagement in the design process. Moreover, \michael{testers} can submit these updated prototypes as ``revisions'' back to designers, providing a more direct means of conveying user needs in specific contexts.\looseness=-1

%% file: sections/sec-030-formative-study.tex
\section{Formative Studies \& Design Goals}
\label{section:formative}

 
\michael{
We began by conducting need-finding interviews with two professional UX designers (D1, D2) from a major technology company, both experienced in designing AI features. We explored their workflows for designing, prototyping, and testing LLM-powered features, as well as the challenges they faced. These discussions inspired the idea of rapidly creating LLM-powered mobile prototypes that can be immediately tested and revised in the wild, leading to the initial development of \systemname. Throughout the development process, we continued to solicit feedback from the original two designers as well as two additional designers (D3, D4), incorporating their insights and suggestions for improvement.
}

Based on the data from these formative studies as well as prior work, we identified three design goals for \systemname:

\begin{itemize}[leftmargin=0.15in]

    \item \textbf{DG1: Rapid and frictionless authoring of medium-fidelity, mobile AI prototypes}. Designers typically prototype AI functionality and UI separately and with different tools. Integrating the AI and UI into a working prototype usually requires significant engineering resources and could often ``take weeks if not months'' (D2) with considerable coordination and communication overhead. 
    While using early versions of \systemname, designers were excited about the possibility of being able to quickly wrap LLM-powered prompts ``in an app shell'' (D3) to create medium-fidelity prototypes. They noted that this would substantially accelerate prototyping velocity and reduce their reliance on engineers, as their goal was \textit{not} to \michael{create high-fidelity, pixel-perfect prototypes or fully-fledged mobile applications}. Three of them were also very interested in the ability to transform ideas into functional prototypes just by describing their ideas in NL.\looseness=-1
    

    \item \michael{\textbf{DG2: Lowering the barriers to on-device testing experience and early-stage feedback}. Designers typically solicit early-stage feedback on AI prototypes by either showing testers static UI mockups, or by showing example model inputs and outputs via prompt editors. When presented with an early version of \systemname, designers found it compelling to potentially ``get the best of both worlds'' (D4)-- a more dynamic and realistic approximation of the AI compared to UI mockups, and a more user-accessible UI compared to the raw model inputs and outputs seen in prompt editors.  Furthermore, designers found it compelling to be able to go from an initial idea to a testable mobile interface quickly, as typical mobile development processes can be painstaking and slow. They were excited to have teammates and other users interact with their prototypes on-the-go, rather than being restricted to desktop environments.}

    \item \textbf{DG3: Expedited design iteration loop}. Traditionally, refining a mobile prototype after user testing is a lengthy and resource-intensive process: designers must make revisions, collaborate with engineers to update the prototype, and schedule new testers, all of which can take days or weeks. To streamline this, designers envisioned a future where the feedback loop is drastically shortened. \michael{Designers were excited about the possibility of receiving tester feedback, then having an instant mechanism to produce an improved version of the prototype for another round of testing. This approach could not only keep testers more engaged and motivated compared to traditional methods, but also reduce designers' dependence on engineers for prototype updates, which typically could take ``days or even weeks'' (D1).}\looseness=-1
\end{itemize}

%% file: sections/sec-040-system.tex
\section{\systemname}

\input{figures/figure-2}

Based on the above design rationales, we describe the system and user interface design of \systemname. 
Our formative studies revealed that designers sought to make prototypes capable of unveiling user intentions, inputs, and reflections on model outputs without being overly complex. Thus, our objective was not to develop a platform for authoring fully-functional mobile applications; instead, we focused on implementing a minimal set of features essential for \michael{designers to explore} the human-AI interactions of the prototypes.

\subsection{Creating Prototypes}

In \systemname, \michael{designers} can build prototypes using three types of configurable UI widgets (Fig. \ref{fig:mobilemaker}):

\begin{itemize}[leftmargin=0.15in]
\item \textbf{Input Widgets} represent different modalities of input that \michael{designers} can provide, including text, dropdown menus (with a list of predefined text values), live-camera photos (captured live from the front or back camera feed on the mobile device), and file uploads (e.g., photos from an on-device album) (Fig. \ref{fig:mobilemaker}A)

\item \textbf{Action Widgets} represent user actions (e.g., a run button or timer) that trigger the prototype to run an LLM prompt with the user's input (Fig. \ref{fig:mobilemaker}B)

\item \textbf{Output Widgets} are used to display the output from generative models, i.e., LLMs, multimodal LLMs, and image generation models (e.g., \cite{noauthor_overview_nodate, noauthor_imagen_nodate})
(Fig. \ref{fig:mobilemaker}C).

\end{itemize}

In addition, output widgets have an editable prompt (Fig. \ref{fig:mobilemaker}D\textsubscript{1}) that is sent to the associated model when generating content. Prompts are composed of three text sections---model instructions, principles, and few-shot style examples--that are concatenated before being sent to a model. 
Prompts may reference the identifier for any input widget in order to have the user's input content from that widget merged into the prompt when calling the model. Output widgets support configuration of model parameters (e.g., temperature) and output parsing behavior (e.g., stop tokens).

Each widget type offers customizable visual elements, such as text input \textit{placeholders} (e.g., ``Enter a color''), widget \textit{labels} (e.g., ``Upload a photo of your room''), and the choice to display model outputs as plain text or in a visual component (e.g., carousel card).
\michael{Designers} can also configure various global parameters, include the prototype's name, description, font style, and layout style, i.e., whether camera controls should be laid out in the list of input controls or as a full screen layout behind all other controls.\looseness=-1

Under the hood, \systemname uses a JSON object to represent the set of widgets and configurations of a prototype (e.g., Fig. \ref{fig:json}), which allows for easy remixing of prototypes (e.g., forking) and revising prototypes with LLMs (sec. \ref{section:nl_revision}).
\systemname renders JSON specifications into interactive prototypes by laying out UI controls for each configured widget (e.g., text input widgets appear as text boxes, photo input widgets appear as a camera interface with a live view from the phone's back camera; Fig. \ref{fig:mobilemaker}F).
On a desktop browser, the prototype is rendered in a mobile phone preview viewport to the right of the configuration toolbars/panels (Fig. \ref{fig:mobilemaker}F); when viewed from a mobile device, the prototype is rendered as a full screen app, omitting the configuration toolbars/panels (Fig. \ref{fig:mobilemaker}G).
Widget modifications are instantly visible from the preview, facilitating rapid iteration throughout the prototyping process.\looseness=-1

\subsection{Prototyping with Natural Language}

From early feedback, we discovered that designers appreciate the capability to quickly convert ideas into functional mobile prototypes (\textbf{DG1})--despite \systemname's straightforward interface, 
some designers found translating high-level concepts to specific low-level widgets cognitively demanding, and could hinder ``spur of the moment'' ideas (D3).
To reduce this barrier, \systemname additionally supports prototyping through two NL features:\looseness=-1

\subsubsection{\michael{Creating with Natural Language}}
\label{section:nl_creation}

In order to allow \michael{designers} to quickly bootstrap a prototype from NL, they can simply provide a brief sketch of their prototype idea (e.g., ``a feature that generates a music playlist from a photo the user takes'').
The system then uses a few-shot LLM prompt 
to instantly generate a prototype. Each example in the few-shot prompt consists of an input NL request and output JSON specification, which encompasses configurations for input, action, and output widgets and as detailed previously. 
Critically, output widget prompts are dynamically generated to reference the corresponding generated input widgets, reducing the need for further manual configuration.\looseness=-1

\subsubsection{\michael{Revising with Natural Language}}
\label{section:nl_revision}

To allow \michael{testers} to rapidly \michael{modify} prototypes (e.g., in response to in-the-wild inspirations), \systemname also supports NL revisions (Fig. \ref{fig:mobilemaker}E).
To achieve this, they can simply provide a brief description of the desired changes \michael{in NL}, such as ``add a dropdown for music genre.''
The system then categorizes the request as either an update to an existing output widget prompt 
or an adjustment to the prototype structure (e.g., adding or removing widgets).
For the former, the system executes a prompt-revision meta-prompt (see appendix \ref{fig:prompt-revision}) 
designed to modify an existing prompt based on \michael{the NL} request.
For the latter, the system combines the \michael{NL request} and the current prototype JSON in a few-shot prompt (similar to the one for NL prototype creation, \michael{where each few-shot example includes a revision request, an initial JSON representation, and an updated JSON representation}),
which outputs a revised JSON specification for the entire prototype.\looseness=-1

When someone requests an NL revision, they will receive a summary outlining the changes between the original prototype and the revised versions. They can toggle between these two versions and decide whether to accept the revision. This feature makes it easy for users to quickly compare and confirm if their requested revision has indeed been implemented \michael{(e.g., during in-the-wild testing)}, enhancing their ability to confidently and efficiently iterate on their design.\looseness=-1

Together, these NL features enable rapid development and require minimal manual configuration, making \systemname's medium-fi prototyping process accessible and approachable to not only designers but also \michael{testers} with limited prototyping or prompting experience.

\input{figures/figure-json}

\subsection{Prototype Testing Experience on Mobile}

\michael{Designers} can share their prototype via URL, which enables team members or other \michael{testers} to try it out in the wild (\textbf{DG2}). When running a prototype on a mobile device, each set of user inputs, model output, and user feedback is saved as a test case. \michael{Designers} can review all collected test cases in a testing dashboard, allowing them to quickly explore use cases and discover unexpected edge cases.

\subsubsection{Dynamic Prototype Revision}

When \michael{testers} experience unexpected outputs or prototype behavior, or they have feedback on how the prototype could be improved to better meet their needs, they can leverage the NL revision feature (sec. \ref{section:nl_revision}) to generate a revised version of the prototype; this can be sent back to the prototype \michael{designer} as a \textit{suggested revision}. 
Each suggested revision includes the \michael{tester's} revision request, the updated JSON prototype specification, and the latest test case generated with the revised prototype prior to submission. 
After revising, \michael{testers} can choose to either continue testing with the modified prototype or revert to the original specification, allowing for both continued iteration on a specific concept and exploration of multiple orthogonal ideas.\looseness=-1

\subsubsection{Prototype Revision Dashboard}
Prototype \michael{designers} can use the revision dashboard (Fig. \ref{fig:revision-dashboard}) to review and examine all proposed revisions, as well as dynamically try out each revision by applying its changes (Fig. \ref{fig:revision-dashboard}A). 
When a revision is applied, \systemname renders the layout and functionality of its JSON prototype specification in the mobile phone preview, then populates the preview with inputs and outputs from the associated test case that the original tester generated and experienced. This enables prototype \michael{designers} to quickly understand and explore the impact of each revision on their prototype, significantly enhancing and accelerating the prototype design loop (\textbf{DG3}).\looseness=-1

\input{figures/figure-revision-dashboard}

The feedback and revision features provide streamlined ways to capture testers' immediate reactions and ideas in the moment, creating a dataset of in situ responses for prototype creators to review during subsequent design iterations.\looseness=-1

%% file: figures/figure-2.tex
\begin{figure*}[t]
    \vspace{-2mm}
    \centering
    \includegraphics[width=0.95\textwidth]{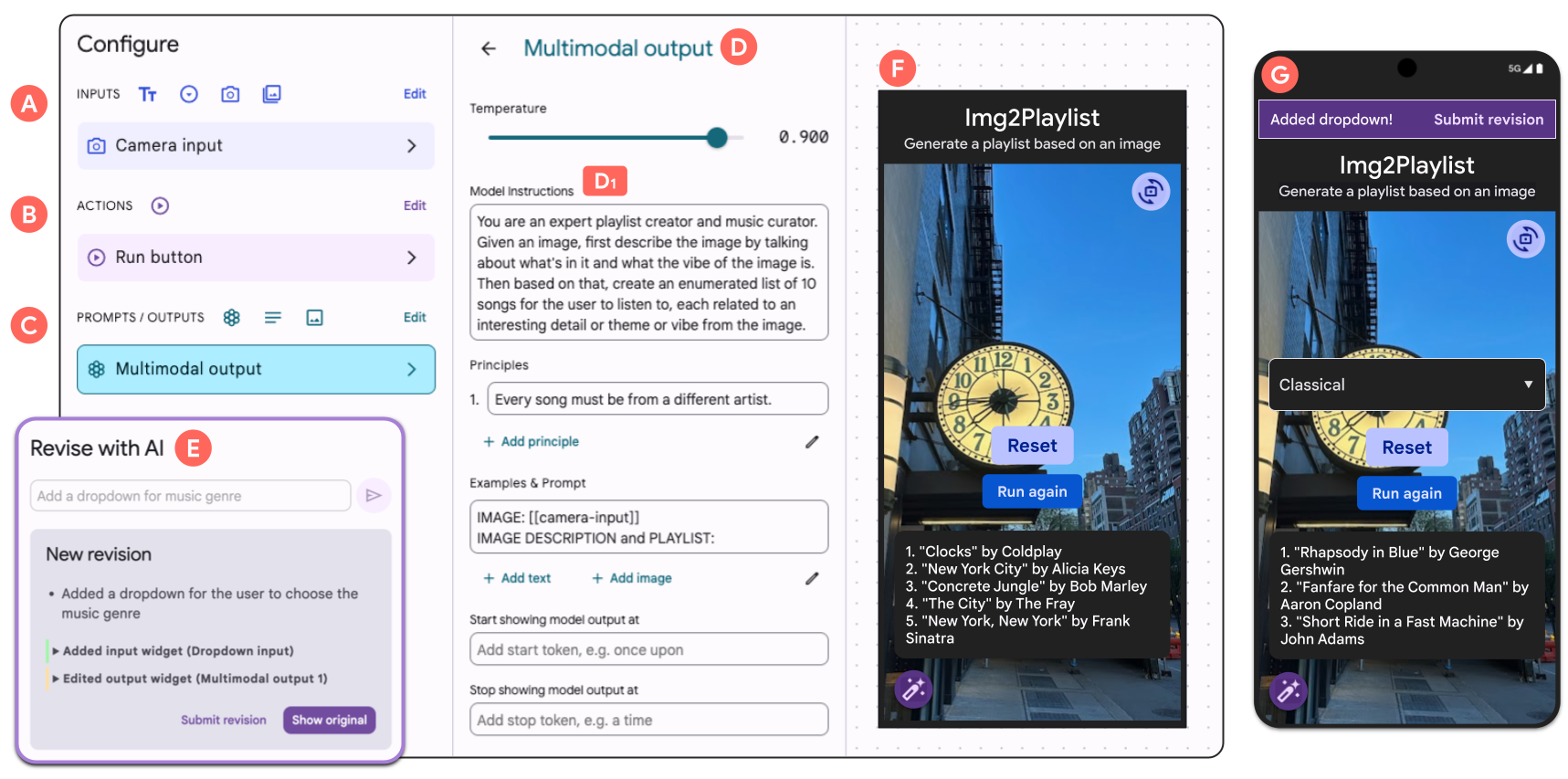}
    \vspace{-2mm}
    \caption{\textbf{\systemname UI}. To build a prototype, users can add input (A), action (B), and output (C) widgets and customize their properties (D), e.g., editing an output widget's prompt (D1). Alternatively, users can create or revise prototypes with natural language via the "Revise with AI" panel (E). Changes are immediately rendered and can be tested in the mobile preview (F). Finally, users can test and revise the prototype on their phones in the wild (G). }
    \label{fig:mobilemaker}
    \vspace{-5mm}
\end{figure*}

%% file: figures/figure-json.tex
\begin{figure}[t]
    \vspace{-1mm}
    \centering
    \includegraphics[width=1.0\linewidth]{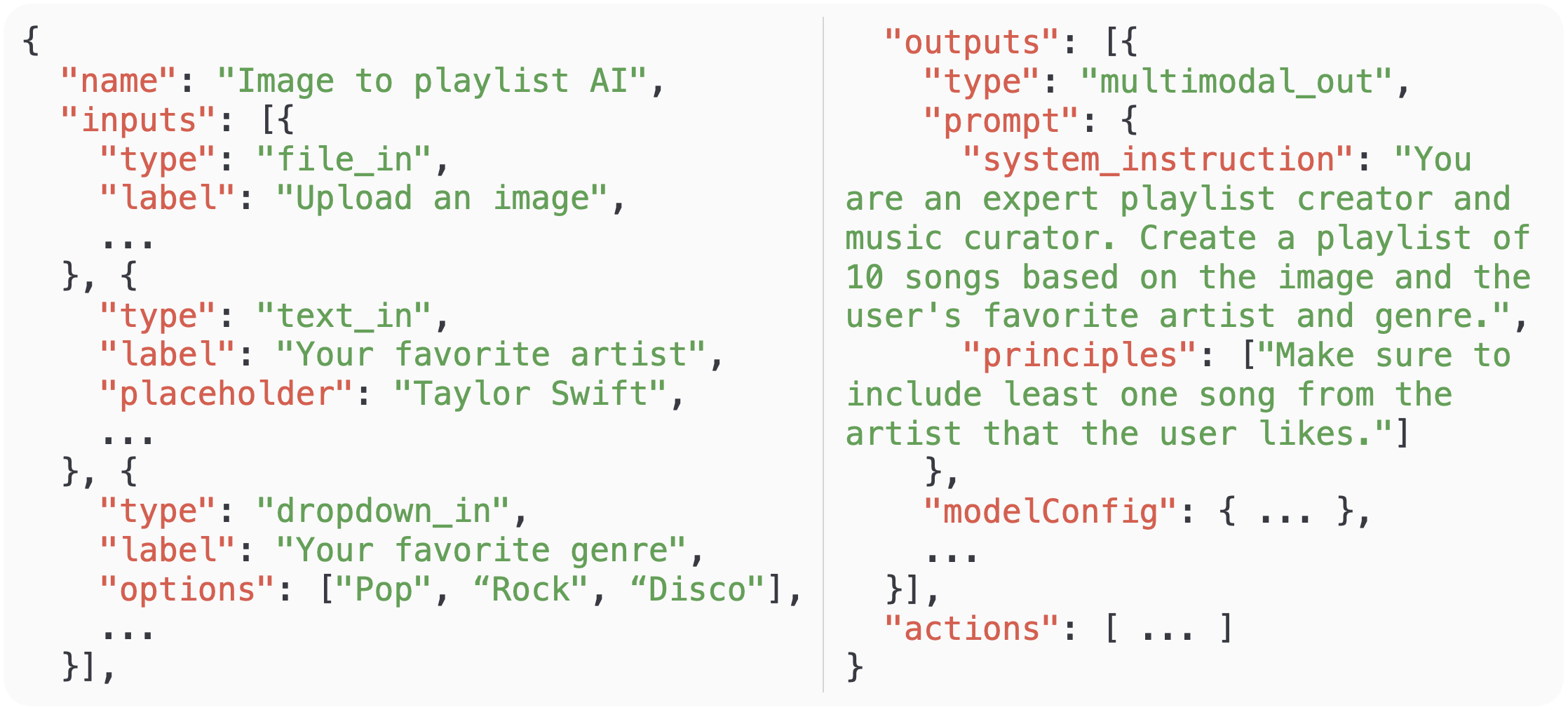}
    \vspace{-6mm}
    \caption{Example prototype JSON representation.}
    \label{fig:json}
    \vspace{-5mm}
\end{figure}




%% file: figures/figure-revision-dashboard.tex
\begin{figure}[t]
\vspace{-1mm}
\centering
\includegraphics[width=0.93\linewidth]{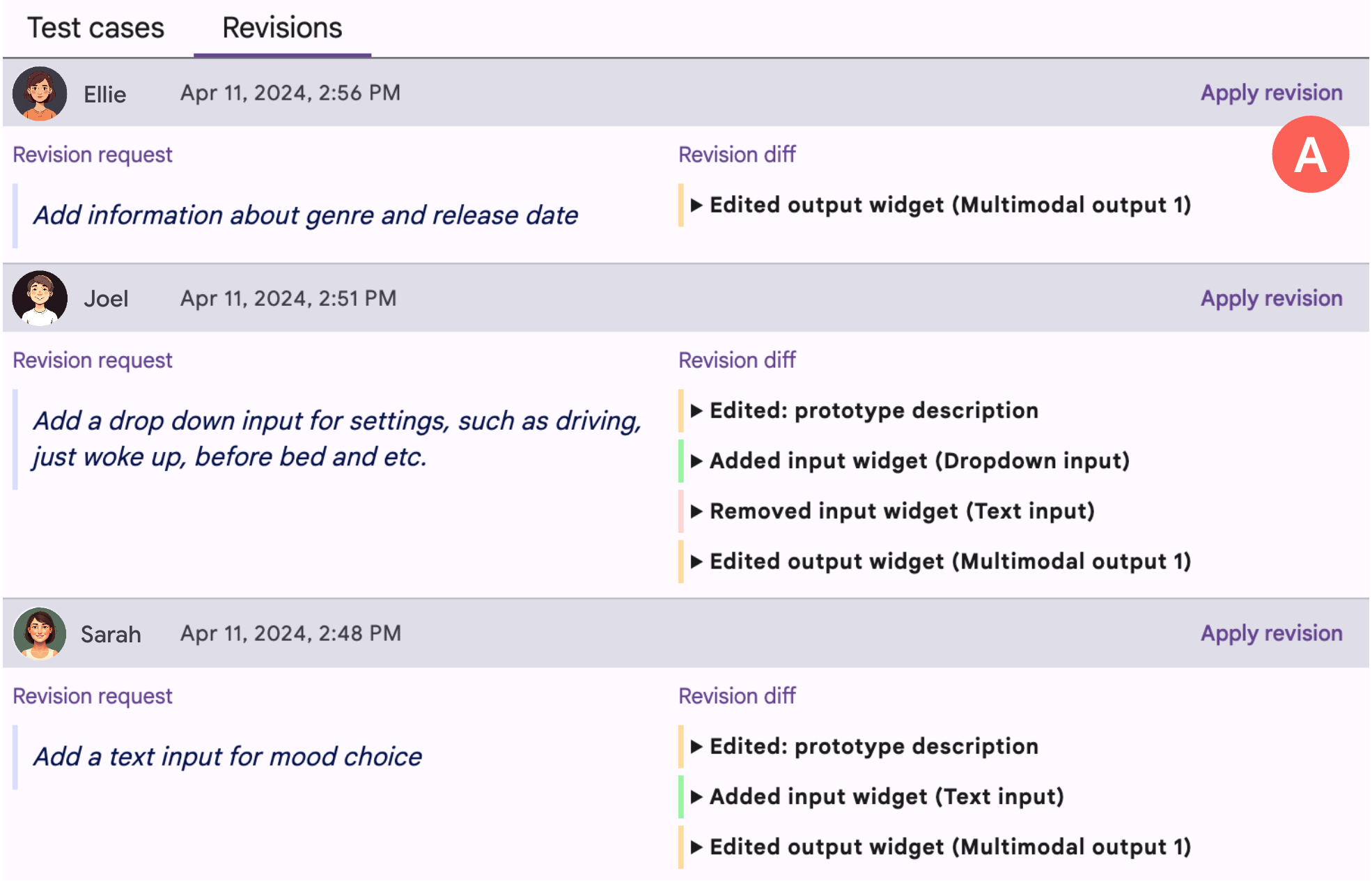}
\vspace{-1mm}
\caption{
\michael{Prototype Revision Dashboard}
}
\label{fig:revision-dashboard}
\vspace{-5mm}
\end{figure}

%% file: sections/sec-050-study.tex
\section{Exploratory User Study}

\savvas{Using \systemname as a probe, we conducted an exploratory study to understand: (1) how \textit{in the wild} tester feedback on functional LLM-powered prototypes compares to status quo tester feedback collected on desktop and (2) how \textit{revising} in the wild affects tester feedback.

The study was within-subjects with two conditions.
In each condition, participants (henceforth referred to as ``testers'') were asked to give feedback on a prototype.
In the \systemname condition, testers gave feedback on a \systemname prototype \textit{in the wild}, and they were also able to \textit{revise} the prototype with the NL revision feature.
In the baseline (desktop) condition, testers clicked through a Figma prototype (with the same UI as the \systemname prototype) that demonstrated the basic interaction.
Afterwards, they searched for images online to test the LLM prompt that would power the prototype in the Google AI Studio prompt editor.
This baseline setup was informed by our formative studies with designers, who indicated that these are typical prototyping workflows their teams use to get early tester feedback on LLM-powered applications.}\looseness=-1


In the \systemname condition, testers used the NL revision feature to revise both the AI and UI functionality while on their phones. In the baseline condition, testers could alter the prompt in a text editor. In both conditions, if testers had trouble altering the prototype or if the NL revision did not work as expected (e.g., the revised prototype did not fit the tester's request), the study facilitators stepped in to help out.


The two prototypes testers experienced were both mobile, multimodal generative AI applications that \savvas{two designers (D1\&2) from our formative study ideated}: (1) Img2Playlist, an application that generates a custom playlist based on an input image, and (2) Img2VideoIdeas, an application that generates ideas for videos to create, also based on an input image.
\savvas{Each tester experienced both prototypes, one per condition and counterbalanced (e.g. P1 experienced Img2Playlist with \systemname and Img2VideoIdeas with the baseline, whereas P4 experienced Img2Playlist with the baseline and Img2VideoIdeas with \systemname).}
We decided on these applications because they target a general audience of users, have the same core interaction pattern, and are actual mobile, generative AI applications professional designers would like to build.\looseness=-1


\subsection{Procedure}

The overall outline for the study is as follows: (1) Testers filled out a consent form prior to the study. (2) During the study, testers spent 50 minutes giving feedback on two different prototypes, one with \systemname (25 minutes) and the other with Figma and a prompt editor (25 minutes), all while thinking aloud. Condition-order and prototype-order were counterbalanced. (3) Testers then completed a post-study questionnaire that compared their experiences giving feedback on each prototype. (4) In a semi-structured interview, testers compared their experiences giving feedback in each condition. 
Each study session took one hour on average.
To situate the task, in both conditions, testers were asked to imagine they were the target user of both of these mobile applications; give any and all feedback they had while interacting with both prototypes; and alter the prototype when they had an applicable revision in mind. In both conditions, if they felt they hit a dead end with their revisions, testers could restart from the beginning, either with the initial prototype (in the \systemname condition) or the initial prompt (in the baseline condition).\looseness=-1

\begin{table}[t]
 \begin{tabular}{p{0.21\linewidth} | p{0.7\linewidth}}
\toprule
\textbf{Measure}          & \textbf{Statement (7-point Likert scale)}                                                                                                                      \\ \midrule
\textbf{Ease}      & 
With \{Setup A/B\}, feedback came easily to me. I was easily stimulated to think of suggestions and feedback for the designer.\\[0.1cm]
\textbf{Communication}        & 
With \{Setup A/B\}, I felt like I could better communicate the changes I wanted to make to the prototype.\\[0.1cm]
\textbf{Actionable Feedback}     & 
With \{Setup A/B\}, I gave feedback that is easily actionable. I think designers can immediately apply my suggestions to their current application. \\[0.1cm]
\textbf{Enjoyable} & 
With \{Setup A/B\}, it was enjoyable to interact with the prototype and provide feedback.\\[0.1cm]
\textbf{Authentic \newline Experience}        & 
With \{Setup A/B\}, I felt like I had a realistic or authentic experience of the envisioned application.                          \\ \bottomrule
\end{tabular}
\vspace{-1mm}
\caption{\textbf{Post-task questionnaire} filled out by testers after they gave feedback on the prototypes from both conditions. 
}
\label{tab:questionnaire}
\vspace{-6mm}
\end{table}

\subsection{Participants}

We recruited 16 participants (6 female, 10 male) from a large technology company, representing diverse professional backgrounds such as designers, product managers, software engineers, UX researchers. We did not target a specific background as the two test applications appeal to a broad audience. Participants were recruited via an email call for participation. Participants were recruited through an email invitation and participated in person in two metropolitan cities in New York City and Pennsylvania. During the study, they explored various environments both within and outside office premises. Each participant received a \$40 gift card for their participation.

\subsection{Questionnaire}
The questionnaire measures (Table \ref{tab:questionnaire}) are derived from established literature on qualities of good  \savvas{\cite{prototyping1,prototyping2,prototyping3_reflective_practitioner}} and have been adapted to fit the perspective of testers.
For instance, one quality of a prototype is its ability to capture the essence of the envisioned application \savvas{\cite{prototyping2}}. 
We thus measured if testers felt they had authentically experienced the envisioned application. Finally, to compare the ratings from the two conditions, we conducted paired sample Wilcoxon tests with full Bonferroni correction, since the study was within-subjects and the questionnaire data was ordinal.\looseness=-1

%% file: sections/sec-060-findings.tex
\section{Findings}

\textit{Quantitative Findings}. The results from the questionnaire are summarized in Figure \ref{fig:study-results-quant}.
Testers reported that giving feedback on the \systemname prototypes was significantly more enjoyable (mean $= 6.12$, $\sigma = 1.11$) than giving feedback on the baseline prototype (mean $= 4.75$, $\sigma = 1.30$, $Z = 10$, $p = .01$), suggesting that 
walking about and experiencing the application in their local environment was a fun and novel experience for testers. 
In addition, testers also found \systemname provided a significantly more authentic experience (mean $= 5.88$, $\sigma = 1.05$) than the baseline prototype (mean $= 3.31$, $\sigma = 1.69$, $Z = 0$, $p <$ .01), suggesting that testers felt that being able to experience the mobile prototype in the wild with real-world inputs provided a more realistic experience of the application. 
Finally, while \systemname was rated higher in ease, communication, and actionable feedback, the differences were not statistically significant. In the following sections, we provide further qualitative context to these results.\looseness=-1

\input{figures/figure-study-result-quant}

\subsection{\savvas{How Testing in the Wild Impacted Tester Feedback}}

Based on the questionnaire results, testers felt that \systemname provided a significantly more authentic experience than the baseline prototype, \savvas{because they could interact with a functional prototype \text{in the wild.}}
\savvas{Experiencing a functional prototype in the wild helped testers:} (1) evaluate if the AI's output was mobile appropriate, (2) experiment with serendipitous edge cases via unconventional, in-the-wild inputs, (3) discover discrepancies between their interpretation of the task and the model's interpretation, and (4) more critically evaluate the model using contextual cues, leading to their identification of rich contextual input in their surroundings that the model did not have access to.
One limitation of in situ testing was that testers couldn't always easily get themselves into representative contexts (e.g., due to scheduling and location constraints), so it wasn't always possible to quickly find the most ``canonical'' inputs (e.g., for \prototypePlaylist, a coffee shop or driving commute picture).\looseness=-1

\subsubsection{\systemname helped testers evaluate if the product concept was mobile appropriate}

While experiencing AI prototypes on \systemname, testers discovered ways in which the prototypes may not be compatible with in-the-wild contexts or form factors. For example, P4 revised \prototypeVideoIdeas to produce a video script along with each video idea, but after trying a few different image inputs, they soon realized that ``generating a full script for the video ideas on the go is overwhelming and too much text.'' Beyond influencing considerations around form factor and mental capacity, the mobile vs. desktop context of use also affected user expectations around the overall ``product fit'' and appropriateness for mobile settings. For example, in the \prototypeVideoIdeas prototype, the model tended to output long-form content ideas, such as how-to videos and history-of videos. Though these may seem acceptable on desktop, P9 (who experienced \prototypeVideoIdeas in the \systemname condition) remarked that they expected more short-form, TikTok-style videos. In contrast, P16 (who experienced \prototypeVideoIdeas in the desktop condition) mentioned that the desktop setting may have prompted them to assume that the prototype was for creators making long-form content. Hence, by enabling testers to experience the prototype authentically on mobile, \systemname enabled testers to not only assess if the look and feel of the model output was mobile-appropriate, but also to reflect on the overall product concept.\looseness=-1


\subsubsection{Experiencing AI prototypes in the wild enabled serendipitous experimentation with edge cases}
\label{edge-cases}

The photos that testers took in the wild tended to be ``noisier'' than on desktop, such as images taken at odd angles, images containing multiple objects of interest, images containing a combination of text and images, etc. The inherent noise in in-the-wild multimedia enabled testers to probe edge cases and boundaries of the model's capabilities and limitations.  While testing \prototypePlaylist with \systemname, P1 noticed a picture of a dog on a nearby desk, with ``Brandy'' written at the edge. Curious to see if the model would pick up on this text, P1 input this photo, and to their surprise, the first song, ``Brandy (You're a Fine Girl)'' by Looking Glass, specifically referenced the text. P1 noted that, with \systemname, ``you are walking around, which allows you to think more about ideas and edge cases,'' whereas on desktop, ``I don't think I would have gotten that [an image with text] just sitting here [using image search].'' Similarly, mobile constraints also led to situations in which the specific content intended to be captured was not always obvious to the model. For instance, P6 took a picture of a far-away cherry blossom, where the photo did not center on the blossom and included other objects in the scene. They were surprised when the model produced a general playlist about New York rather than the cherry blossoms. Overall,  encountering these edge-case scenarios helped testers probe the boundaries of the model's capabilities and limitations.

\subsubsection{Unconventional inputs uncovered discrepancies between the AI's and user's interpretation of the problem}
\label{discrepancies}

In addition to encountering model edge cases (e.g., images with unique compositions), testers were also more likely to encounter objects and settings that were unique or unconventional given the \textit{use case} (e.g., for \prototypePlaylist: pictures of hairspray, carton of milk, blank walls, etc.). This spontaneity and serendipity helped them uncover multiple possible interpretations of the problem. For example, while interacting with \prototypePlaylist on mobile, P2 input an image of a carton of milk (Figure \ref{fig:input_examples}b), expecting a playlist with a ``coffee shop vibe,'' but instead received a more literal response: songs with titles related to milk (e.g., ''Cream'' by Prince). These spontaneous experiences uncovered multiple interpretations of the problem: it is unclear if a ``relevant'' playlist means that it ought to be relevant to the literal object in the picture, to the vibe of the scene, or to the look of an artist's album cover.\looseness=-1

Meanwhile, on desktop, testers tended to search for conventional images that more often yielded expected model outputs. For example, P8 searched for ``John Legend'' in an image search engine and selected an image of one of his canonical album covers. They expected \prototypePlaylist to produce a playlist with songs only from that album, and it mostly did. This targeted approach lacked the serendipity of the \systemname inputs. As P8 explained, ``\textit{If I do a search on Google, I have an intention already... there's no randomness, and randomness is useful.}'' It's possible that image search results may be less likely to expose intent mismatches between testers and the AI, because they themselves were found through being highly linked to a user's explicit text query. Overall, the images testers sourced in the baseline condition were more intent-driven and less ambiguous, which led to fewer discoveries of model output discrepancies.\looseness=-1

\subsubsection{Contextual cues enabled a more informed evaluation of the model's outputs}
\label{contextual}
When experiencing prototypes on \systemname, testers had access to a wealth of sensory, spatial, and temporal context surrounding their image inputs, including the sounds and scents present in the environment, the encompassing landscape surrounding the image, and the events that occurred before and after the image was taken. This additional context helped them more critically evaluate model outputs that they would have otherwise deemed reasonable in the baseline condition. For example, P7 took a picture of two coworkers playing a relaxed game of pool in the early afternoon (Figure \ref{fig:input_examples}b), when most people had just finished eating lunch. They were surprised when the model erroneously identified the scene as ``competitive.'' P7 realized that it may be difficult for a model to accurately assess the vibe with just a single, static image. ``\textit{On the phone, I was in the environment in which the photo was taken... I have more context, not just what's captured in that one-frame, but what's also before or after. It makes me more critical about the result.}'' Similarly, P4 (while indoors) wished the model could take into consideration the fact that it was raining just outside their window, evoking a solemn mood. Overall, testing the prototype in-situ not only enabled testers to assess the model's outputs through the surrounding context it was situated in, but also helped them identify contextual cues the model did \textit{not} have access to. 


\subsection{How NL Revision Changed AI Prototyping Practices}
Prompted by these in-the-wild discrepancies, testers revised the prototypes, ranging from \textit{adding UI controls} (e.g. a dropdown menu for music genres or an input textbox to specify the audience for the video ideas) to \textit{changing model behavior} (e.g. generating a description that associates each song in the playlist to the input image, or outputting a video script to accompany each idea).
NL revision enabled testers to play an active, fulfilling role in the design process, and it also shortened the test-feedback loop.
\savvas{However, revising also potentially limited tester feedback to smaller, incremental revisions of the application's functionality.
Whereas with Figma, testers gave higher level feature requests and ideas.
This suggests that perhaps the feedback these prototypes provide are complementary and useful at different stages of the design process.}\looseness=-1



\subsubsection{Revising enabled testers to provide better guardrails to the AI and reconsider the prototype's interaction pattern}

When the model behavior did not match user intent, testers leveraged the NL revision tool to revise the prototype, such as by adding UI controls. 
As described in sec \ref{discrepancies}, P2 discovered that the model sometimes generated playlists based on the literal object (e.g., ``Cream'' by Prince) rather than the vibe of the scene. When confronted with this issue, P8 revised their prototype to add a dropdown menu that allows testers to select whether the model should generate playlists based on the ``vibe'' or the ``object'' shown in the image. In subsequent testing, they found that having this input helped guardrail the model from solely providing playlists that were literally related to the objects in the image. Similarly, when P6 took a far-away picture of cherry blossoms, and the model did not pick up on the cherry blossom in the cluttered image (described in sec \ref{edge-cases}), P6 added a textbox input widget so that testers can specify which object they wanted the playlist to be based on.
After experimenting with this revision, P6 started to prefer having the playlist generated by the text.
They appreciated the flexibility starting with text provided, and ultimately, they questioned the interaction pattern of starting with an image.
Overall, the NL revision feature enabled testers to proactively react to model discrepancies, as well as critically reflect on the core interaction pattern (e.g., input and output types) of the application.\looseness=-1

\input{figures/figure-3}

\subsubsection{NL revise enabled testers to critically evaluate their own feedback in situ}
\label{evaluate_on_feedback}
With \systemname's NL revision feature, testers could immediately ``live'' and experience their own feedback within the updated prototype, which often helped them rethink their initial feedback. P4 initially revised \prototypeVideoIdeas to ``Add a script for each video idea'', but after testing the revised prototype, they realized that viewing a video script on mobile was overwhelming. They felt this realization could have only occurred on \systemname: ``Walking around and trying it on mobile gave me a much better intuition on what I liked and disliked... for example, generating a full script for the video ideas on the go is overwhelming and too much text. I would have probably said I wanted that feature on desktop without realizing this. That realization is much more likely to happen if you're actually using it, as opposed to being asked to think about it.'' Beyond this, NL revision also empowered testers to make subsequent, ``cascading'' revisions (or cascading feedback based on a revision). In the example above, when P8 added the dropdown menu for selection of ``vibe'' or ``object'', they found that the model actually had difficulty assessing vibe from one image alone (described in sec \ref{contextual}).\looseness=-1

\subsubsection{Revising helped shorten the test-feedback loop in \systemname}

Revising in \systemname helped testers feel like they were actively and iteratively improving the prototype. While experiencing \prototypePlaylist, P2 added a dropdown menu for genre to control the types of songs output by the application and then added a description to accompany each song to explain their relevance to the input image. They felt that these additions tangibly improved the base application by introducing greater user control of the playlist and tightening the connection between the input image and output playlist: ``In the mobile prototype, I felt like we were stepping up the stairs to get to a better and better, more refined version of the north-star version of the application.'' In the baseline, even though their feedback was used to alter the prompt in the editor, P2 felt they were giving feedback on the prompt's output, but the application as a whole was not improving. Echoing these thoughts, P16 described how experiencing their suggested revisions felt more active and fulfilling: ``\textit{You can complete this loop of giving feedback and seeing it in action. This experience is definitely much better, rather than describing your frustrations, and hopefully getting them fixed.}'' Testers' revisions in \systemname would build on each other, which helped them feel (1) more active in the design process and (2) that they were tangibly improving the application.\looseness=-1

\subsubsection{Testers felt restricted to give feedback and revisions enabled by \systemname}

While testers enjoyed immediately experiencing their feature suggestions, many also felt restricted by the revisions that were possible in \systemname. P6 explained, ``\textit{With [\systemname], you're a bit limited by what you can add to the UI. And whatever revision you add will be a bit more permanent in the mobile prototype.}'' With \systemname, testers tended to focus on feedback that could lead to new revisions they could experience, e.g., adding a dropdown or a text input to steer the output playlist, potentially at the expense of providing feedback that could not be immediately rendered as a revision (e.g., thoughts on alternative user journeys or workflows).
\savvas{Meanwhile, in the baseline condition, testers often ideated more ``out there'' feature requests and entirely different functionalities with Figma.
For example, P14 suggested that Img2VideoIdeas could be embedded into existing social media platforms and altered to also support caption generation for video posts.
This higher level, more ``out there'' feedback elicited through Figma might be more useful to designers during early stage ideation, whereas the more granular and grounded feedback enabled by revising might be more useful further along the design process, when designers are refining their application's functionality.
Overall, because they elicit quite different feedback, Figma and MobileMaker might be employed at different parts of the design process, depending on the needs of the designer.
}


\subsubsection{Challenges in implementing revisions}

While we intended for the testers to be able to independently update the prototypes, in practice, the facilitators occasionally needed to step in to help.
In general, the LLM prompt powering revisions worked best when testers (1) input a single revision and (2) clearly specified that revision.
\savvas{For example, revision requests that added a single input like ``add a dropdown for a music genre'' or made a simple, clear adjustment to the model's output, e.g. ``give me 5 songs instead of 10'' tended to work well immediately.}
However, on a few occasions, testers entered revision requests with multiple changes (e.g., add two separate dropdowns and remove a text input) and the prompt failed to produce all of the specified changes.
This type of error was mostly mitigated by educating testers on the feature during our short demonstration. One more challenging issue was clearly delineating requests. While P13 was testing \prototypeVideoIdeas, they wanted the application to take in multiple images that would be used together to inform three video ideas. They wrote: ``I want three photos to be considered together,'' but the model interpreted this to mean creating a separate video idea for each image.  If P13 had better specified their request, it likely would have been implemented correctly; however, testers often wanted to provide a quick description and have the model correctly extrapolate their intent.

%% file: figures/figure-study-result-quant.tex
\begin{figure}[t]
\vspace{-2mm}
\centering
\includegraphics[width=0.92\linewidth]{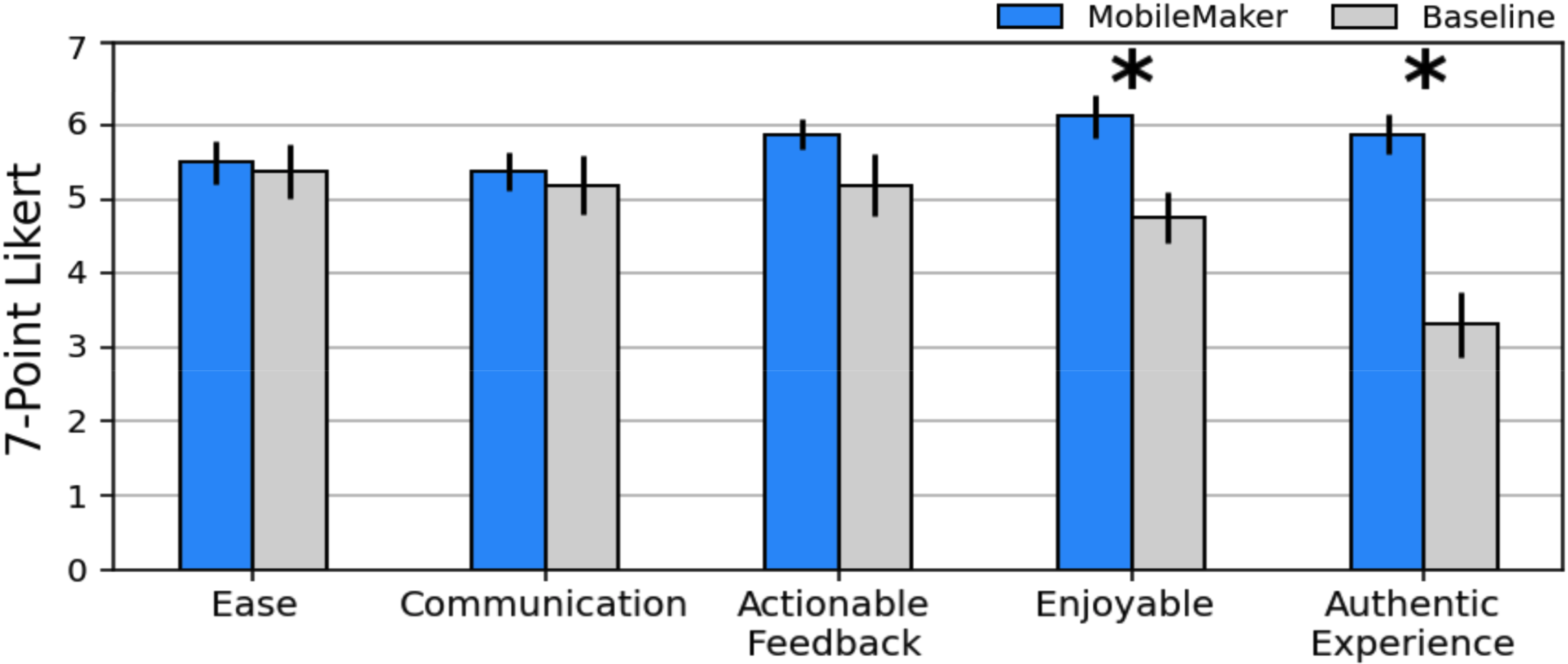}
\vspace{-2mm}
\caption{Questionnaire results comparing the two conditions. Bars are standard error and an asterisk indicates a statistically significant difference (after full Bonferroni correction).}
\label{fig:study-results-quant}
\vspace{-5mm}
\end{figure}

%% file: figures/figure-3.tex
\begin{figure*}[t]
    \vspace{-2mm}
    \hspace{-19mm}
    \centering
    \begin{subfigure}[b]{0.4\textwidth}
        \centering
        \setlength{\tabcolsep}{2pt}
        \begin{tabular}{rcccc}
            \multirow{2}{12mm}{\vspace{-2mm}\parbox{14mm}{\small(a)\newline Example desktop inputs}} &
            \includegraphics[width=0.23\linewidth]{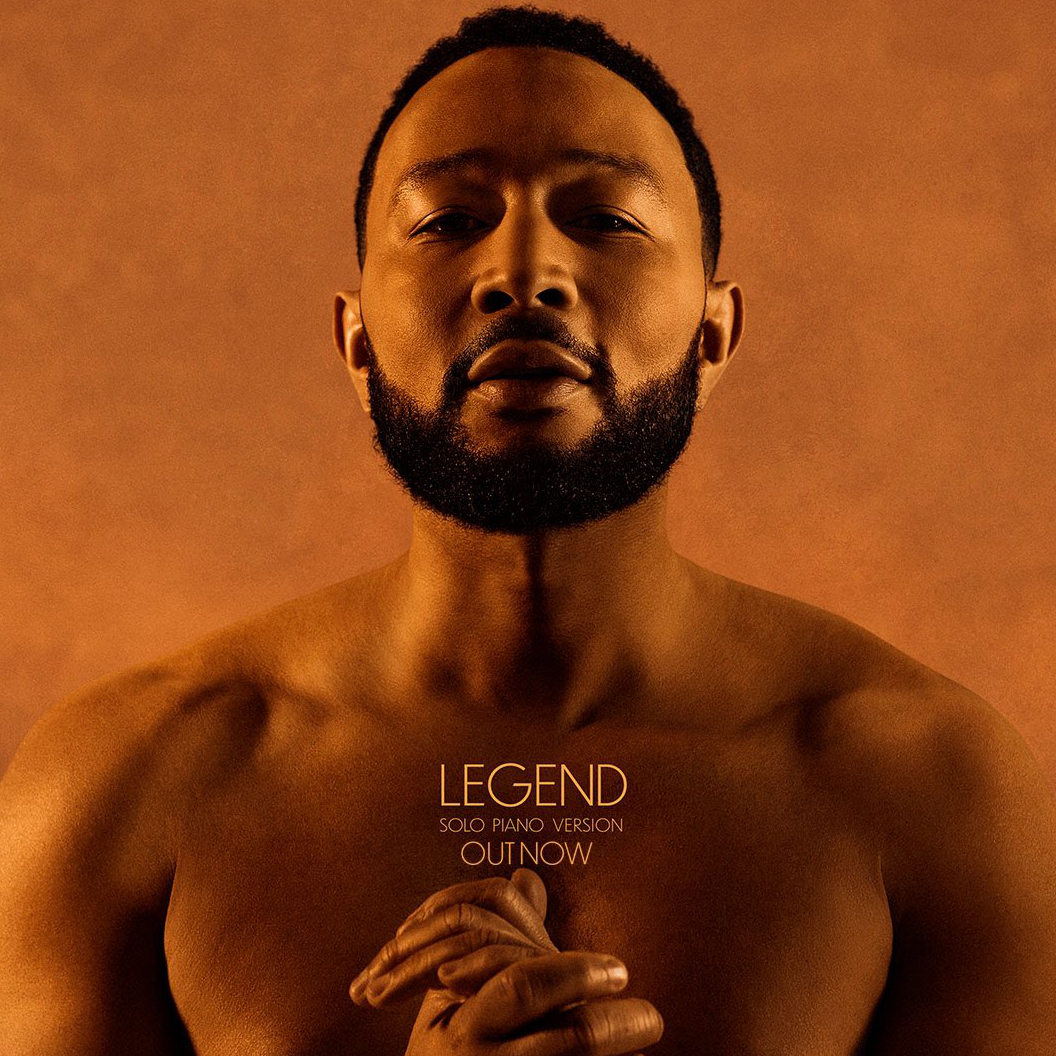} &
            \includegraphics[width=0.23\linewidth]{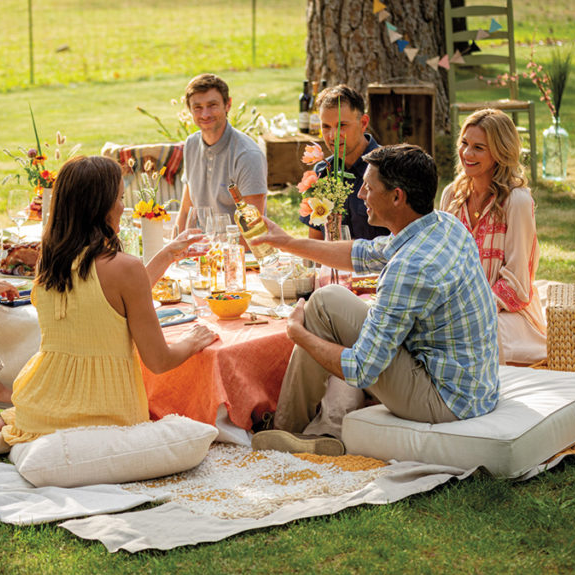} &
            \includegraphics[width=0.23\linewidth]{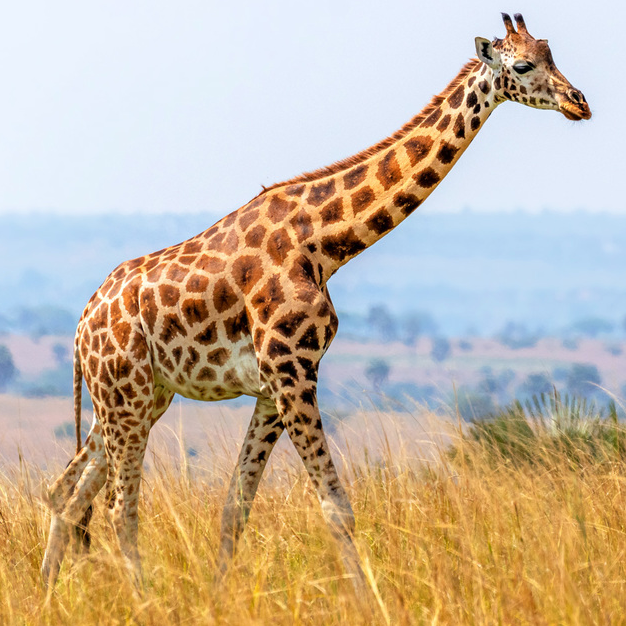} &
            \includegraphics[width=0.23\linewidth]{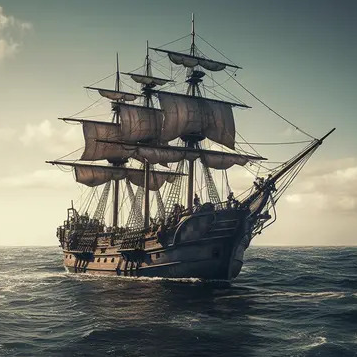} \\
            ~ &
            \includegraphics[width=0.23\linewidth]{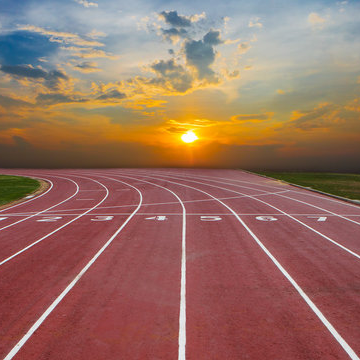} &
            \includegraphics[width=0.23\linewidth]{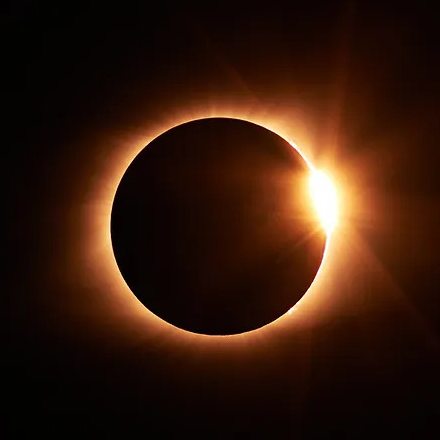} &
            \includegraphics[width=0.23\linewidth]{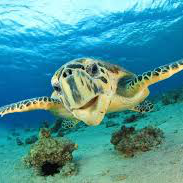} &
            \includegraphics[width=0.23\linewidth]{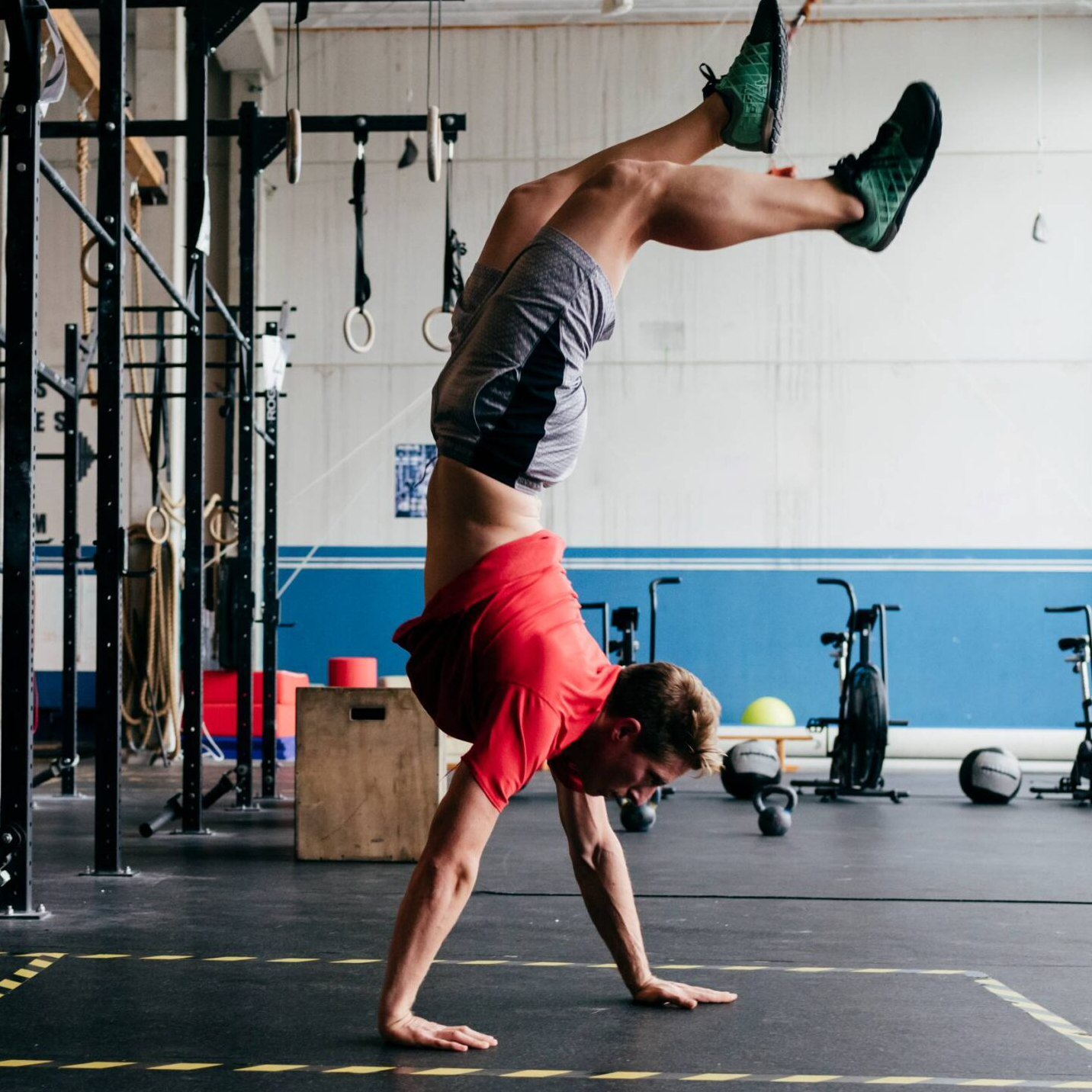} 
        \end{tabular}
    \end{subfigure}%
    \hspace{18mm}
    \begin{subfigure}[b]{0.4\textwidth}
        \centering
        \setlength{\tabcolsep}{2pt}
        \begin{tabular}{rcccc}
        \multirow{2}{15mm}{\vspace{-2mm}\parbox{18mm}{\small(b)\newline Example in-the-wild inputs}} &
            \includegraphics[width=0.23\linewidth]{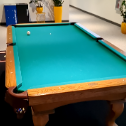} & 
            \includegraphics[width=0.23\linewidth]{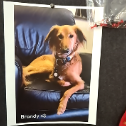} & 
            \includegraphics[width=0.23\linewidth]{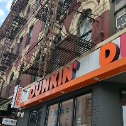} &
            \includegraphics[width=0.23\linewidth]{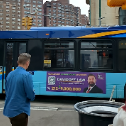} \\
            ~ &
            \includegraphics[width=0.23\linewidth]{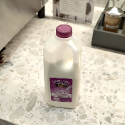} & 
            \includegraphics[width=0.23\linewidth]{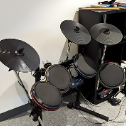} &
            \includegraphics[width=0.23\linewidth]{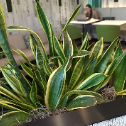} & 
            \includegraphics[width=0.23\linewidth]{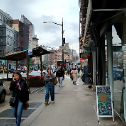}  
        \end{tabular}
    \end{subfigure}%
    \vspace{-2mm}
    \caption{Examples of test images tried by study participants during the desktop (a) and in-the-wild (b) conditions.}
    \label{fig:input_examples}
    \vspace{-6mm}
\end{figure*}

%% file: sections/sec-070-design-reflection.tex
\section{Design Reflection Exercise}\label{sec:design-reflection-exercise}

In addition, we conducted an initial probe on how designers might interact with \michael{testers}' prototype revisions. From the \systemname condition of the previous study, we presented a dashboard of participants' suggested revisions, as well as their revised prototypes, to the two designers who had initially prototyped \prototypePlaylist and \prototypeVideoIdeas. Designers spent around 30 minutes perusing the dashboard and applying the suggested revisions while thinking aloud.\looseness=-1

D1 remarked that, relative to traditional forms of user feedback, experiencing users' dynamically-runnable revised prototypes was ``very powerful and a more convincing form of feedback.'' They thought it made user feedback more tangible and compelling compared to vague requests such as ``I want [feature request],'' which are easier for designers to neglect or dismiss in practice. Beyond this, both designers found it crucial to have access to the \textit{in-the-wild context} embedded within those revised prototypes. In \systemname, designers could run the revised prototype on the actual inputs experienced by the user (e.g., picture of pool table), and see the outputs the user saw. Those test cases also served as tangible evidence for suggested revisions. For example, the designers were initially perplexed by a sudden request to change the original \prototypePlaylist feature to ``video2playlist.'' After noticing that the participant had previously inputted a series of photos depicting different scenes in a game room, they hypothesized that the participant preferred video over still images to better capture the dynamic context and ambiance. In the future, designers would like to see more of the ``user journey'' and reasoning behind the proposed revisions, so they could better understand users' motivations while running the prototypes. \looseness=-1

Designers wished that there was an easier way to synthesize the different threads of feedback that participants provided in \systemname so that they could be more effectively utilized. One proposed solution by D2 was for \systemname to automatically organize the suggested revisions into distinct themes, such as ``providing rationale for song choices in the output,'' ``using video instead of image as input,'' or ``add more UI controls to influence the type of playlist,'' as well as statistics on their frequency. As one designer explained, ``it's not just one person reading through [the feedback] but a whole team that needs to be convinced that this needs to happen in this way,'' so it would be important to have sufficient evidence to help prioritize and facilitate more targeted design adjustments.

%% file: sections/sec-080-discussion.tex
 \section{Discussion}

\subsection{
Agile AI-Prototyping Practices in the Age of LLMs}

Short design iteration loops coupled with frequent testing have long been advocated for in the design and software development communities. 
This paper expands on the question of what might be possible during the early phases of AI design if (1) designers can quickly push functional AI prototypes out into the wild and 
(2) testers can contribute both feedback and updated functional prototypes while in the field.
In light of this, we may witness a parallel shift in the roles and responsibilities of designers, user researchers, developers, and end-users. 
Designers and developers might begin to borrow practices from user research (e.g., in-the-field user observations); 
likewise, user researchers may take on more designerly and creative roles by troubleshooting or refining user-proposed revisions directly in the field.
Overall, in situ prototyping could introduce a blurring of boundaries between these traditional roles, broadening participation of AI development to include more diverse professions and end-users. 
While this may inevitably introduce new frictions (e.g., increased coordination costs of disambiguating roles, potential suboptimal designs by end-users \cite{norman_design_2013}, etc.), future work could also capitalize on the new opportunities brought on by this blurring of roles, e.g., designers and developers more deeply considering off-the-desktop user needs; UX researchers taking a more active role in implementing user feedback; and end-user communities personalizing and configuring AI to their own needs.\looseness=-1

\subsection{Making In Situ Medium-fi Prototypes more ``Sketchable''}

Beyond giving passive feedback, \systemname users were empowered to actively participate in the prototyping process by revising prototypes and immediately ``living'' them in the wild. 
On one hand, this enabled users to more critically assess the feedback they gave (see sec \ref{evaluate_on_feedback}, where a participant suggested a revision, and realized after testing it that it was inappropriate for mobile). 
On the other hand, this introduced the potential risk that users might inadvertently ``short circuit'' the design process by prematurely discarding ideas that are not immediately feasible to implement within \systemname.
Future work could address this by emphasizing to users that they are exploring initial feature ideas instead of final product implementation.
One approach might involve modifying the UI of \systemname-built prototypes to resemble a sketch or low-fidelity paper prototype.
Additionally, when running the NL revision feature, \systemname could produce multiple variations of the revision, e.g., three different video scripts varying in length and structure, 
to reinforce the concept that these are preliminary explorations with multiple potential implementations.\looseness=-1

\subsection{Uncovering Design Axes and Personas}

While designers found the aggregated feedback and revisions useful for understanding user needs, they also expressed a desire to see the revisions clustered thematically so it would be easier to digest (see sec \ref{sec:design-reflection-exercise}).
Synthesizing this feedback at a larger scale, with perhaps a hundred users, could lead to a few interesting possibilities, such as uncovering key personas or revealing different conflicting perspectives on the future direction of the envisioned application. 
For example, some users of \prototypePlaylist might be ``music experts'' who want fine-grained controls to steer the playlist (e.g., through descriptors like ``ethereal'' or specific sub-genres); others might prefer a simpler UI with fewer controls so they can listen to music as soon as possible. 
These different perspectives could define a broad design space for \prototypePlaylist applications. 
In addition, future work could examine how LLMs might be leveraged to help designers transform aggregated feedback into a well-defined design space, both to better understand different user personas and product scope, and to identify under-explored regions within the design space. By uncovering early design axes, product scoping requirements, and personas for consideration, in situ prototyping could shift user feedback even further \textit{upstream} in the AI development process.

\subsection{Exploring Prototype Variants Based on Contexts}

With the ability to instantly transform a piece of feedback or feature idea into an improved version of the prototype in situ, participants in the user study tended to engage in a ``depth-first search'' pattern, where they meticulously added or refined features through successive iterations. 
While thorough, this method risks leading testers down a narrow path, resulting in a form of tunnel vision where alternative ideas or unexpected user needs might be overlooked. 
To mitigate this, \systemname could encourage more breadth-first, ``parallel-prototyping'' \cite{dow_parallel_2010} early in the testing process, such as leveraging the advanced reasoning capabilities of LLMs \cite{brown_language_2020} to proactively suggest prototype variants based on both the existing prototype JSON configuration and the user's context (e.g., location, weather conditions). For a recommendation feature that suggests outfits based on a photo of the user's wardrobe and a selfie, \systemname could generate variants that let users provide additional input about their preferred style (e.g., casual, sporty, formal), or consider current weather and season. Beyond reducing design fixation, this approach might also encourage in-the-field testers to consider diverse conditions and assist them with prototyping AI that resonates with users' immediate and situational needs.

\subsection{Limitations and Future Work}
During the \systemname portion of the study, participants mostly explored areas inside or adjacent to their office buildings. 
To enrich results, \michael{participants} could ideally use the applications in a variety of natural settings, such as near their homes or at favorite spots in the city.
However, for practicality and to observe \michael{participants} as they experienced the \systemname prototypes in person, we kept the locations standard across participants.
\savvas{Also, future work can examine how MobileMaker impacts tester feedback on a wider variety of applications.}
Finally, while \systemname's feature set is not exhaustive, it enabled us to probe the value and opportunities of in situ AI prototyping and testing in our exploratory study. In the future, \systemname can be expanded to support building prototypes with multi-screen, \michael{multi-LLM calls} workflows and a wider array of UI inputs and outputs.\looseness=-1


%% file: sections/sec-090-conclusion.tex
\section{Conclusion}
To explore the opportunities for in situ AI prototyping and testing, we developed \systemname, with which designers can (1) rapidly create mobile AI prototypes and (2) enable testers to authentically experience the prototype in the wild and revise it in real-time.
In our exploratory study, we found that by testing in the wild, testers were able to serendipitously experiment with image edge cases.
In addition, NL revision enabled testers to critically evaluate their own feedback but also potentially limited the feedback they gave to what was actionable in \systemname.
Future work might explore further supporting (1) users with providing feedback on in situ medium fidelity prototypes, as well as (2) designers with deriving insights from this feedback at scale.

%% file: appendix/prompt-creation.tex
\begin{figure*}
    \begin{tcolorbox}
    \begin{ttfamily}
    \begin{scriptsize}
        You are an expert at turning simple app ideas into working mobile application prototype configurations. Prototypes consist of a series of input widgets and output widgets. Input widgets are UI controls that allow the user to provide different types of input. Here is a list of the available input widget types:

        1. TEXT: Text fields allowing the user to input text.
        
        2. CAMERA: The mobile device camera, allowing the user to take photos.
        
        3. UPLOAD\_IMAGE: An upload control that allows the user to upload an image.
        
        4. OPTIONS\_LIST: A dropdown control with a list of fixed text options the user may choose from. Prefer this input type when the input is text but needs to be constrained to several options. Input widgets must have an empty options list unless they are of type OPTIONS\_LIST.
        
        \medskip
        
        Output widgets describe different types of output that can be generated by the prototype by a generative model and the model instructions, principles, and prompt that are sent to the model in order to generate that output. The model instructions tell model what it is supposed to do, the principles describe rules for how the model should generate and format it's output, and the prompt combines the inputs and prompts the model to generate the output. Here is a list of the available output widget types:

        1. TEXT: The model generates text output. Only valid if all inputs are of type TEXT.
        
        2. MULTIMODAL: The model generates text output, but takes text and image inputs. Only valid if the intended output to generate is text.
        
        3. IMAGE\_GENERATION: The model generates image output. Only valid if the intended output to generate is an image instead of text.
        
        \medskip
        
        Your task is the following: given an idea for an app prototype, generate a prototype configuration for that app. Try to be creative when coming up with the combination of input modalities. Follow these guidelines when generating the prototype:
        
        1. Output prompts may **ONLY** reference inputs. They may not reference other outputs!
        
        2. All inputs must be referenced by one or more of the output widgets!
        
        3. Names for prototypes should be short (three words of fewer) and fun, and should capture people's attention when the app is released on app stores.
        
        4. Before making the prototype, you should first think about how the new app should work as a whole from an app developer's perspective (i.e., starting the thought with "This xxxx app helps users with xxxx..."). Update the "functionalDescription" with this thought.
        
        5. Do NOT repeat content in the model\_instructions and the prompt.
        
        \medskip
        
        Here are a few examples:
        
        IDEA: "I want an app that helps users pick an outfit on any given day."
        
        \medskip
        
        PROTOTYPE CONFIG:
        \{
          "appInfo": \{
            "funName": "Style Me!",
            "shortDescription": "Help me pick an outfit!",
            "functionalDescription": "This fashion app recommends outfits based on what a user have in their closet, the weather as well as the user's style. The app will accept a photo of the user's closet showing all the clothes that they own, the current weather condition outside, as well as their style, and it will output a list of clothing items that they can wear that day."
          \},
          "inputs": [
            \{
              "id": "input-01-camera",
              "type": "CAMERA",
              "description": "A photo of your closet showing your clothes",
              "options": []
        
            \},
            \{
              "id": "input-02-text",
              "type": "TEXT",
              "description": "Today's weather",
              "options": []
            \},
            \{
              "id":"input-03-text",
              "type": "TEXT",
              "description": "Your style",
              "options": []
            \}
          ],
          "actions": [
            \{
              "id": "action-01-run",
              "type": "RUN\_BUTTON"
            \}
          ],
          "outputs": [
            \{
              "id": "output-01-multimodal",
              "type": "MULTIMODAL",
              "description": "A list of outfit that you can wear",
              "modelInstructions": "Generate an outfit suggestion based on the items in the picture of the user's closet, the current weather, and the user's style.",
              "principles": ["Provide a list of three outfit options"],
              "prompt": "CLOSET PHOTO: [[input-01-camera]]{\textbackslash}nWEATHER: [[input-02-text]]{\textbackslash}nUSER'S STYLE: [[input-03-text]]{\textbackslash}nOUTFIT OPTIONS:",
              "triggeredBy": "action-01-run"
            \}
          ]
        \}
        
        \medskip
        
        IDEA: "A harry potter sorting hat application that tells the user which house they are in"
        
        \medskip
        
        PROTOTYPE CONFIG:
        \{
          "appInfo": \{
            "funName": "Hogwarts Sorting Hat",
            "shortDescription": "See what Hogwarts House you belong to!",
            "functionalDescription": "This Sorting Hat app uses a photo of you and your listed values and desires to determine which Hogwarts house you belong to."
          \},
          "inputs": [
            \{
              "id": "input-01-text",
              "type": "TEXT",
              "description": "Your values and desires",
              "options": []
            \},
            \{
              "id": "input-02-camera",
              "type": "CAMERA",
              "description": "A photo of you",
              "options": []
            \}
          ],
          "actions": [
            \{
              "id": "action-01-run",
              "type": "RUN\_BUTTON"
            \}
          ],
          "outputs": [
            \{
              "id": "output-01-multimodal",
              "type": "MULTIMODAL",
              "description": "Your determined house",
              "modelInstructions": "You are the Sorting Hat from Harry Potter. Given the photo of the user and text describing their given values, determine what house they are in.",
              "principles": [],
              "prompt": "PHOTO: [[input-02-camera]]{\textbackslash}nVALUES: [[input-01-text]]{\textbackslash}nHOUSE:",
              "triggeredBy": "action-01-run"
            \}
          ]
        \}
        
        \medskip
        
        IDEA: "Make an app that helps me decide where to put my new furniture"
        
        \medskip
        
        PROTOTYPE CONFIG:
        \{
          "appInfo": \{
            "funName": "Furniture Placer",
            "shortDescription": "Visualize your space with new furniture",
            "functionalDescription": "This app combines a photo of your space with a photo of the furniture you want to buy to help you visualize the furniture in your space."
          \},
          "inputs": [
            \{
              "id": "input-01-camera",
              "type": "CAMERA",
              "description": "A photo of the room",
              "options": []
            \},
            \{
              "id": "input-02-upload-image",
              "type": "UPLOAD\_IMAGE",
              "description": "An image of the new furniture that you're considering buying",
              "options": []
            \}
          ],
          "actions": [
            \{
              "id": "action-01-run",
              "type": "RUN\_BUTTON"
            \}
          ],
          "outputs": [
            \{
              "id": "output-01-image-generation",
              "type": "IMAGE\_GENERATION",
              "description": "An image of the room with the furniture in the ideal place",
              "modelInstructions": "Given a photo of a room and the new furniture that the user wants to buy, generate an image showing the furniture placed in the ideal location in the room.",
              "principles": [],
              "prompt": "ROOM PHOTO: [[input-01-camera]]{\textbackslash}nNEW FURNITURE: [[input-02-upload-image]]",
              "triggeredBy": "action-01-run"
            \}
          ]
        \}
        
        \medskip
        
        IDEA: "\colorbox{black}{\color{white}\textbf{PLACEHOLDER FOR USER INPUT}}"
        
        \medskip
        
        PROTOTYPE CONFIG:
    \end{scriptsize}
    \end{ttfamily}
    \end{tcolorbox}
    \caption{Natural language prototype creation prompt}
    \label{fig:proto_creation}
\end{figure*}

%% file: appendix/prompt-classifier.tex
\begin{figure*}[!ht]
    \begin{tcolorbox}
    \begin{ttfamily}
    \begin{scriptsize}
        You are an expert at **making revisions** to mobile app prototypes. Given a user request, determine if fulfilling that request requires adding/editing/removing input and output widgets from the prototype (op\_type: REVISE\_PROTOTYPE\_STRUCTURE) or changing the principles that control the style and/or format of the models' output for a prototype (op\_type: REVISE\_PRINCIPLES).

        \medskip
        
        REQUEST: "Also allow the user to provide their favorite color"
        
        RESULT: \{ "thought": "The user wants to add a new input field for their favorite color", "op\_type": "REVISE\_PROTOTYPE\_STRUCTURE" \}

        \medskip
        
        REQUEST: "Make sure the output is a bulleted list"
        
        RESULT: \{ "thought": "The user wants to control the formatting of the model output", "op\_type": "REVISE\_PRINCIPLES" \}

        \medskip
        
        REQUEST: "Change the purpose of the app so that it generates an image instead of just text"
        
        RESULT: \{ "thought": "The user wants to change the purpose of the app and that requires different outputs to be generated", "op\_type": "REVISE\_PROTOTYPE\_STRUCTURE" \}

        \medskip
        
        REQUEST: "Keep the response short. No more than 5 words"
        
        RESULT: \{ "thought": "The user wants the model output to be terse", "op\_type": "REVISE\_PRINCIPLES" \}

        \medskip
        
        REQUEST: "\colorbox{black}{\color{white}\textbf{PLACEHOLDER FOR USER INPUT}}"
        
        RESULT: 
    \end{scriptsize}
    \end{ttfamily}
    \end{tcolorbox}
    \caption{Natural language request classifier}
    \label{fig:request-classifier}
\end{figure*}

%% file: appendix/prompt-revision.tex
\begin{figure*}[!ht]
    \begin{tcolorbox}
    \begin{ttfamily}
    \begin{scriptsize}
        Given a request from a user and a list of principles, determine the operation that should be conducted to update the principles to fulfill the user's request.

        \medskip

        REQUEST: "Make sure to output 3 ideas in the list"
        
        PRINCIPLES: \{ "principles": ["1. Generate a list of video ideas based on the video input", "2. Each list item should start with the name of the movie"] \}
        
        OPERATION: \{ "thought": "The user wants to add a new principle indicating that the list must include exactly 3 ideas.", "op\_type": "ADD\_TO\_PROMPT", "op": \{ "principle": "The list must include exactly 3 ideas", index: null \} \}

        \medskip
        
        REQUEST: "Make sure the story is at least two paragraphs long"
        
        PRINCIPLES: \{ "principles": ["1. Generate a short story based on the user's inputs", "2. The story needs at least one named character", "3. The story must be one paragraph long"] \}
        
        OPERATION: \{ "thought": "The user wants to revise the principle at index 1 to say that the story must be at least two paragraphs long instead of one paragraph long.", "op\_type": "REVISE\_PROMPT", "op": \{ "principle": "The story must be at least two paragraphs long", index: 2 \} \}

        \medskip
        
        REQUEST: "Every song should be from a different artist"
        
        PRINCIPLES: \{ "principles": ["1. Given an image, identify a minimum of 4 interesting details in the image and then create a playlist with 10 songs with a rational relating to an interesting detail or theme from the image.", "2. Consider the artistic style and aesthetic of the image and share how they relate to musical moods and themes.", "3. Relate the image and moods to a musical style or genre.", "4. If there is a person present in the frame, consider their clothes, accessories, mood, and actions.", "5. Generate a playlist title with a creative and funny voice that is quirky and a little ironic."] \}
        
        OPERATION: \{ "thought": "The user wants to add a new principle indicating that each of the recommended songs must be from a unique artist", "op\_type": "ADD\_TO\_PROMPT", "op": \{ "principle": "Ensure that each recommended song is from a different artist", index: null \} \}

        \medskip
        
        REQUEST: "Output the list in markdown format"
        
        PRINCIPLES: \{ "principles": ["1. Generate a list of fun clown names for the user based on their picture and name", "2. Names must be funny", "3. Names should be short"] \}
        
        OPERATION: \{ "thought": "The user wants to add a new principle indicating that the list must be output in the markdown format.", "op\_type": "ADD\_TO\_PROMPT", "op": \{ "principle": "The list must be output in the markdown format", index: null \} \}

        \medskip
        
        REQUEST: "Don't output movie names. Just generate the script ideas"
        
        PRINCIPLES: \{ "principles": ["1. Generate a list of video ideas based on the video input", "2. Each list item should start with the name of the movie"] \}
        
        OPERATION: \{ "thought": "The user wants to remove the principle at index 1 that indicates the list should include movie names.", "op\_type": "REMOVE\_FROM\_PROMPT", "op": \{ "index": 1 \} \}

        \medskip
        
        REQUEST: "\colorbox{black}{\color{white}\textbf{PLACEHOLDER FOR USER INPUT}}"
        
        PRINCIPLES: \colorbox{black}{\color{white}\textbf{PLACEHOLDER FOR CURRENT PROMPT PRINCIPLES}}
        
        OPERATION: \
    \end{scriptsize}
    \end{ttfamily}
    \end{tcolorbox}
    \caption{Prompt revision prompt}
    \label{fig:prompt-revision}
\end{figure*}

%% file: appendix/structure-revision.tex
\begin{figure*}[!ht]
    \begin{tcolorbox}
    \begin{ttfamily}
    \begin{scriptsize}
        You are an expert at **making revisions** to mobile app prototypes. Prototypes
        consist of a series of input widgets and output widgets. Input widgets are UI
        controls that allow the user to provide different types of input. Here is a list
        of the available input widget types:
        
        \medskip
        
        1. TEXT: Text fields allowing the user to input text.
        
        2. CAMERA: The mobile device camera, allowing the user to take photos.
        
        3. UPLOAD\_IMAGE: An upload control that allows the user to upload an image.
        
        4. OPTIONS\_LIST: A dropdown control with a list of fixed text options the user may 
        choose from. Prefer this input type when the input is text but needs to be constrained 
        to several options. Input widgets must have an empty options list unless they are 
        of type OPTIONS\_LIST.
        
        \medskip
        
        Output widgets describe different types of output that can be generated by the 
        prototype by a generative model and the model instructions, principles, and prompt 
        that are sent to the model in order to generate that output. The model instructions tell 
        model what it is supposed to do, the principles describe rules for how the model 
        should generate and format it's output, and the prompt combines the inputs and prompts 
        the model to generate the output. Here is a list of the available output widget types:
        
        \medskip
        
        1. TEXT: The model generates text output. Only valid if all inputs are of type TEXT.
        
        2. MULTIMODAL: The model generates text output, but takes text and image inputs. 
        Only valid if the intended output to generate is text.
        
        3. IMAGE\_GENERATION: The model generates image output. Only valid if the intended 
        output to generate is an image instead of text.
        
        \medskip
        
        Your task is the following: given the initial prototype config and a request for 
        modification, think of one possible revision to the prototype. The revision 
        can add new inputs or outputs, modifying existing inputs or outputs, or  remove 
        unnecessary inputs or outputs based on the user's modification request.
        
        \medskip
        
        Follow these guidelines when generating the revised prototype:
        
        \medskip
        
        1. Output prompts may **ONLY** reference inputs. They may not reference other outputs!
        
        2. DO NOT remove or edit any existing principles for an output widget! Simply copy 
        the existing principles to the output. You may suggest new principles by adding 
        them to the end of the principles list. If you create a new output, you may add 
        principles to the principles list. You may edit the model instructions and the prompt.
        
        3. The model instructions do not need to repeat any content that is already in 
        the principles.
        
        4. All inputs must be referenced by one or more of the output widgets!
        
        5. Names for prototypes should be short (three words of fewer) and fun, and 
        should capture people's attention when the app is released on app stores.
        
        6. If you need to add a new TEXT output, first try to see if you can reuse or 
        repurpose existing TEXT outputs before adding a new one. Please note that you 
        should also make corresponding updates to the prompt based on the modification 
        to the input and output elements.
        
        7. Before making the revision, you should first think about how the new app should 
        work as a whole (rather than specifically what changed) after the revision from 
        an app developer's perspective (i.e., starting the thought with "This xxxx app 
        helps users with xxxx..."). Update the "functionalDescription" with this thought. 
        Also provide a short and concise list summarizing what changed compared to the original 
        setup that's written in an active voice and present tense, such as "Add xxx and 
        xxx as additional inputs", "Change from xxxx to xxxx", "Remove xxx output", "Make 
        it clear the output should be in xxx format / mention xxx". Add this list of 
        revisions to the output under the key "summaryOfChanges".
        
        \medskip
        
        PROTOTYPE CONFIG:
        \{
          "appInfo": \{
            "funName": "Style Me!",
            "shortDescription": "Help me pick an outfit!",
            "functionalDescription": "This fashion app recommends outfits based on what the user has in their closet."
          \},
          "inputs": [
            \{
              "id": "input-01-camera",
              "type": "CAMERA",
              "description": "A photo of your closet showing your clothes",
              "options": []
            \}
          ],
          "actions": [
            \{
              "id": "action-01-run",
              "type": "RUN\_BUTTON"
            \}
          ],
          "outputs": [
            \{
              "id": "output-01-multimodal",
              "type": "MULTIMODAL",
              "description": "An outfit suggestion",
              "modelInstructions": "Generate an outfit suggestion based on the items in the picture of the user's closet.",
              "principles": ["Ensure the suggestion is a complete outfit including a top, bottom, shoes, and accessories."],
              "prompt": "CLOSET: [[input-01-camera]]{\textbackslash}nOUTFIT SUGGESTION:",
              "triggeredBy": "action-01-run"
            \}
          ]
        \}
        
        \medskip
        
        USER MODIFICATION REQUEST: "allow users to also input today's weather and their preferred style"
        
        \medskip
        
        REVISED PROTOTYPE CONFIG:
        \{
          "appInfo": \{
            "funName": "Style Me!",
            "shortDescription": "Help me pick an outfit!",
            "functionalDescription": "This fashion app recommends outfits based on what a user have in their closet, the weather as well as the user's style. The app will accept a photo of the user's closet showing all the clothes that they own, the current weather condition outside, as well as their style, and it will output a list of clothing items that they can wear that day."
          \},
          "summaryOfChanges": ["Add today's weather and the user's style as additional inputs"],
          "inputs": [
            \{
              "id": "input-01-camera",
              "type": "CAMERA",
              "description": "A photo of your closet showing your clothes",
              "options": []
            \},
            \{
              "id": "input-02-text",
              "type": "TEXT",
              "description": "Today's weather",
              "options": []
            \},
            \{
              "id": "input-03-text",
              "type": "TEXT",
              "description": "Your style",
              "options": []
            \}
          ],
          "actions": [
            \{
              "id": "action-01-run",
              "type": "RUN\_BUTTON"
            \}
          ],
          "outputs": [
            \{
              "id": "output-01-multimodal",
              "type": "MULTIMODAL",
              "description": "A list of outfit that you can wear",
              "modelInstructions": "Generate an outfit suggestion based on the items in the picture of the user's closet, the current weather, and the user's style.",
              "principles": ["Ensure the suggestion is a complete outfit including a top, bottom, shoes, and accessories."],
              "prompt": "CLOSET: [[input-01-camera]]{\textbackslash}nCURRENT WEATHER: [[input-02-text]]{\textbackslash}nUSERS STYLE: [[input-03-text]]{\textbackslash}nOUTFIT SUGGESTION:",
              "triggeredBy": "action-01-run"
            \}
          ]
        \}
        
    \end{scriptsize}
    \end{ttfamily}
    \end{tcolorbox}
    \caption{Prototype structure revision prompt}
\end{figure*}
\begin{figure*}[!t]\ContinuedFloat
    \begin{tcolorbox}
    \begin{ttfamily}
    \begin{scriptsize}
        PROTOTYPE CONFIG:
        \{
          "appInfo": \{
            "funName": "Furniture Placer",
            "shortDescription": "Where to put my furniture?",
            "functionalDescription": "This app helps users visualize new furniture in their space."
          \},
          "inputs": [
            \{
              "id": "input-01-camera",
              "type": "CAMERA",
              "description": "A photo of the room",
              "options": []
            \},
            \{
              "id": "input-02-upload-image",
              "type": "UPLOAD\_IMAGE",
              "description": "An image of the new furniture that you're considering buying",
              "options": []
            \}
          ],
          "actions": [
            \{
              "id": "action-01-run",
              "type": "RUN\_BUTTON"
            \}
          ],
          "outputs": [
            \{
              "id": "output-01-multimodal",
              "type": "MULTIMODAL",
              "description": "The ideal place to place the furniture",
              "modelInstructions": "You are an expert interior designer. Given a photo of a room, and the new furniture that the user wants to buy, describe the ideal place in the room to arrange the furniture.",
              "principles": ["Ensure the description references the location in the room pictured in the photo."],
              "prompt": "ROOM PHOTO: [[input-01-camera]]{\textbackslash}nNEW FURNITURE: [[input-02-upload-image]]{\textbackslash}nSUGGESTED ARRANGEMENT:",
              "triggeredBy": "action-01-run"
            \}
          ]
        \}
        
        \medskip
        
        USER MODIFICATION REQUEST:  "Also show an image of the new furniture in the room"
        
        \medskip
        
        REVISED PROTOTYPE CONFIG:
        \{
          "appInfo": \{
            "funName": "Home Decor and Travel",
            "shortDescription": "Decorate your home and book your travel at the same time",
            "functionalDescription": "This home decor application can not only provide furniture placement suggestions but also allow users to book hotel stays based on the furniture layout of the hotel rooms. The app lets a user take a photo of their room and upload an image of their new furniture. Additionally, it asks the users for the city that they'll travel to and the dates of travel. It then recommends the ideal spot to place the new furniture, generate an image of the room with the furniture in the ideal place, and provide a list of hotels in the destination city that the user is traveling to during the dates provided with a similar furniture layout."
          \},
          "summaryOfChanges": ["Add the destination city and travel dates as additional inputs", "Emphasize that the output should also provide hotel recommendations"],
          "inputs": [
            \{
              "id": "input-01-camera",
              "type": "CAMERA",
              "description": "A photo of the room",
              "options": []
            \},
            \{
              "id": "input-02-upload-image",
              "type": "UPLOAD\_IMAGE",
              "description": "An image of the new furniture that you're considering buying",
              "options": []
            \}
          ],
          "actions": [
            \{
              "id": "action-01-run",
              "type": "RUN\_BUTTON"
            \}
          ],
          "outputs": [
            \{
              "id": "output-01-multimodal",
              "type": "MULTIMODAL",
              "description": "The ideal place to place the furniture",
              "modelInstructions": "You are an expert interior designer. Given a photo of a room, and the new furniture that the user wants to buy, describe the ideal place in the room to arrange the furniture.",
              "principles": ["Ensure the description references the location in the room pictured in the photo."],
              "prompt": "ROOM PHOTO: [[input-01-camera]]{\textbackslash}nNEW FURNITURE: [[input-02-upload-image]]{\textbackslash}nSUGGESTED ARRANGEMENT:",
              "triggeredBy": "action-01-run"
            \},
            \{
              "id": "output-02-image-generation",
              "type": "IMAGE\_GENERATION",
              "description": "A generated image of the room with the furniture in the ideal place",
              "modelInstructions": "You are an expert interior designer. Given a photo of a room, and the new furniture that the user wants to buy, generate an image showing what the furniture will look like in their room.",
              "principles": [],
              "prompt": "ROOM PHOTO: [[input-01-camera]]{\textbackslash}nNEW FURNITURE: [[input-02-upload-image]]{\textbackslash}nFURNITURE ARRANGED IN ROOM:",
              "triggeredBy": "action-01-run"
            \}
          ]
        \}
        
        \medskip
        
        PROTOTYPE CONFIG:
        \colorbox{black}{\color{white}\textbf{PLACEHOLDER FOR CURRENT PROTOTYPE JSON}}
        
        \medskip
        
        USER MODIFICATION REQUEST:  "\colorbox{black}{\color{white}\textbf{PLACEHOLDER FOR USER INPUT}}"
        
        \medskip
        
        REVISED PROTOTYPE CONFIG:
    \end{scriptsize}
    \end{ttfamily}
    \end{tcolorbox}
    \caption{Prototype structure revision prompt (cont.)}
    \label{fig:structure-prompt}
\end{figure*}